\documentclass[conference]{IEEEtran}
\usepackage{booktabs}

\IEEEoverridecommandlockouts
\usepackage{cite}
\usepackage[caption=false]{subfig}
\usepackage{amsmath,amssymb,amsfonts}
\usepackage{algorithm}
\usepackage{algpseudocode}
\usepackage{amsmath} 
\usepackage[hidelinks]{hyperref}
\usepackage{graphicx}
\usepackage{textcomp}
\usepackage{xcolor}
\usepackage{placeins}
\usepackage{graphicx}
\usepackage{dsfont}
\usepackage{enumitem}
\def\BibTeX{{\rm B\kern-.05em{\sc i\kern-.025em b}\kern-.08em
    T\kern-.1667em\lower.7ex\hbox{E}\kern-.125emX}}
\begin{document}

\title{Communication-Semantic-Aware RDMA Loss Recovery for QP-scalable Hyperscale AI Training}
\author{
\IEEEauthorblockN{
Xiaoxiang Hua\IEEEauthorrefmark{1},
Tong Zhang\IEEEauthorrefmark{1},
Zhenjiang Dong\IEEEauthorrefmark{2},
Kun Zhu\IEEEauthorrefmark{1},
Jiakun Bao\IEEEauthorrefmark{3},
and Fengyuan Ren\IEEEauthorrefmark{4}
}
\IEEEauthorblockA{
\IEEEauthorrefmark{1}Nanjing University of Aeronautics and Astronautics, Nanjing, China\\
\IEEEauthorrefmark{2}School of Computer Science, Nanjing University of Posts and Telecommunications, Nanjing, China\\
\IEEEauthorrefmark{3}China Tower Co., Ltd. Jiangsu Branch, Nanjing, China\\
\IEEEauthorrefmark{4}Tsinghua University, Beijing, China\\
Email: \{huaxiaoxiang, zhangt, zhukun\}@nuaa.edu.cn, dongzhenjiang@njupt.edu.cn, baojk10@163.com, renfy@tsinghua.edu.cn
}
}

\maketitle

\begin{abstract}
Current artificial intelligence (AI) infrastructures widely adopt Remote Direct Memory Access (RDMA) to support high-performance communication. Training trillion-parameter models involves frequent collective communication operations, such as All-Reduce and All-to-All, which generate intensive RDMA traffic. Existing RDMA deployments predominantly use the reliable connection (RC) model, where each process pair requires a dedicated queue pair (QP). This leads to poor scalability: since the RDMA-capable network interface card (RNIC) can cache only a few thousand QPs, and excess entries trigger PCIe round-trip penalties. Meanwhile, global synchronization makes training sensitive to tail latency, where a few packet loss can delay iteration completion. To address these challenges, we propose Communication-Semantic-Aware Unreliable Datagram (CSA-UD), a novel RDMA loss recovery mechanism that combines scalability and reliability. CSA-UD decouples data transmission from loss recovery and dynamically adjusts the loss detection interval, accelerating tail recovery and exploiting the synchronization semantics of distributed training. It further supports multipath transmission and bitmap-guided reassembly, enabling high throughput without requiring lossless fabrics. Testbed experiments and ns-3 simulations show that CSA-UD significantly reduces tail latency under large-scale collective  communication. Under high network load, it achieves better scalability than RC and over 30\% lower 99th percentile flow completion times compared to with counterparts.
\end{abstract}

\section{Introduction}
The scale of modern AI training has entered the trillion-parameter era, placing unprecedented demands on the underlying compute infrastructure. To support such dense model training, intelligent computing centers have rapidly evolved as the backbone of next-generation AI systems. It is now common for a single cluster to deploy tens of thousands to over one hundred thousand GPUs or AI accelerators. Some platforms are even planning training systems with more than 200,000 devices to meet the increasing model complexity and data throughput requirements~\cite{zhao2025deepseekv3, pilz2025_trends,li2024understanding}.

Modern intelligent computing centers clusters are equipped with increasingly powerful multicore servers. These machines are typically provisioned with 128\,GB to 2\,TB of DRAM and interconnected through high-bandwidth, low-latency fabrics such as InfiniBand, Omni-Path, or RoCE, all of which support Remote Direct Memory Access (RDMA)~\cite{tran2024ohio}. By enabling zero-copy and kernel bypass, RDMA significantly reduces end-to-end latency and boosts throughput. As a result, it has been widely adopted for large model training~\cite{liu2023janus}. To fully leverage distributed computing resources, large-scale training workloads partition trillion-parameter models and datasets across nodes, performing collective communication operations such as all-reduce or all-to-all over RDMA to synchronize gradients and intermediate results in each iteration~\cite{li2024understanding}. These workloads exhibit four key communication characteristics:
\begin{enumerate}
    \item \textbf{Rich connectivity}: Each process exchanges data with every other process, forming a fully connected graph.
    \item \textbf{High burstiness}: Millions of point-to-point messages are injected into the network during short synchronization windows.
    \item \textbf{Sensitivity to tail latency}: The progress of each iteration is regulated by the slowest flow (tail flow), which determines the training time per iteration~\cite{gangidi2024rdma_meta}.
    \item \textbf{Requirement for reliability}: As each message often carries a gradient or model update essential to global model correctness, even a single lost packet can compromise convergence or necessitate costly retransmissions.
\end{enumerate}

RDMA communication is orchestrated through QPs, each of serving as a dedicated virtual channel between communicating peers. In most deployments, QPs are instantiated in the reliable connection (RC) mode. Although RC performs well at small scales, it faces severe scalability bottlenecks in large training jobs: each pair of communicating processes maintains a dedicated QP, resulting in $O(N^2)$ QP states.

This QP explosion quickly exhausts the on-chip static random-access memory (SRAM) of RNICs. For instance, Mellanox ConnectX-4/5 cards provide only $2\,\mathrm{MB}$ of SRAM to store address mappings and QP metadata. Since each RC QP consumes roughly $375$ bytes, only around $5{,}000$ QPs can be cached~\cite{monga2021flock, singhvi2020rma, wang2024optimized}. Although ConnectX-6 claims to support $131{,}072$ QPs, empirical measurements show that throughput drops beyond $256$ concurrent QPs due to cache misses and PCIe round-trips~\cite{monga2021flock, singhvi2020rma}.

In addition to state bloat, RC’s single-path data transmission easily causes congestion. Congestion control schemes such as DCQCN~\cite{erannSIGCOMM2015DCQCN} and TIMELY~\cite{mittal2015timely} offer partial relief, but cannot fully prevent losses. To avoid packet loss, operators often deploy Priority Flow Control (PFC), which introduces head-of-line (HOL) blocking, and network-wide congestion that undermines the low-latency advantage of RDMA.

Several solutions attempt to alleviate these issues. On-NIC schemes like XRC~\cite{xrc-ibta} and DCT~\cite{rosenbaum2018dct} reduce per-connection state but do not eliminate state bloat~\cite{nvidia2020prm}. More aggressive approaches such as IRN\cite{mittal2018revisiting}, SRNIC~\cite{srnic-nsdi23}, and csRNA~\cite{ma2022survey} attempt to virtualize QPs or extend RNIC state, but incur significant hardware modifications and deployment complexity. An alternative strategy is to offload reliability to the host. For instance, ERD~\cite{wang2024optimized} uses Unreliable Datagram (UD) transport and implements a software-based NACK retransmission mechanism, enabling thousands of peers to be served by a small QP pool. However, ERD suffers from three critical limitations: (1) a fixed loss-scan interval delays detection of tail losses, prolonging iteration duration; (2) TCP-based NACK signaling incurs kernel context switches; and (3) packet reordering requires data copying to large buffers, breaking the zero-copy principle of RDMA.

To overcome these limitations, we propose Communication-Semantic-Aware Unreliable Datagram (CSA-UD), a scalable RDMA loss recovery mechanism  tailored for hyperscale AI training. CSA-UD adopts the UD mode to avoid  per-peer QP overhead associated with RC. Each process maintains a small, fixed number of QPs to communicate with all peers, fundamentally resolving QP scalability. Besides, CSA-UD is co-designed with the collective communication semantics of AI training and focuses on minimizing tail latency. To this end, CSA-UD adopts  an \textit{adaptive loss detection mechanism} that dynamically adjusts the loss detection interval based on the remaining volume of unacknowledged data and network conditions. CSA-UD enhances reordering through a bitmap-guided buffer mapping structure. Since CSA-UD operates entirely in userspace without requiring any modifications to RNIC hardware, it offers high deployability for future exascale AI infrastructure.

We evaluate CSA-UD using both testbed experiments and ns-3 simulations. Compared to RC, CSA-UD reduces QP usage, avoids PCIe-induced cache thrashing, and mitigates congestion by supporting multipath transmission. Compared to other UD counterparts, CSA-UD further reduces 99th percentile latency  by up to 30\%.

\section{Background and Motivation}

\subsection{RDMA Basics}
RDMA  is built upon a transport abstraction known as QP, which defines the data transmission channel between two communicating endpoints. Each QP consists of a Send Queue (SQ), a Receive Queue (RQ), and a Completion Queue (CQ). Prior to data transmission, both sender and receiver must independently create and initialize their local QPs and exchange required metadata.

Architecturally, QP metadata—including current states, MR mappings, and WQE pointers—is both maintained in host memory and cached in the RNIC’s on-chip SRAM for fast access. In high-concurrency scenarios, the limited RNIC cache capacity can be quickly exhausted. When active QPs exceed the cache limit, cache misses force the RNIC to retrieve QP metadata  via PCIe from host memory, significantly increasing access latency, which is  a major source of latency amplification, especially for tail flows in collective communication.
\subsection{RDMA Reliability Mechanisms }
\begin{figure}[t]
\centering
\includegraphics[width=0.48\textwidth]{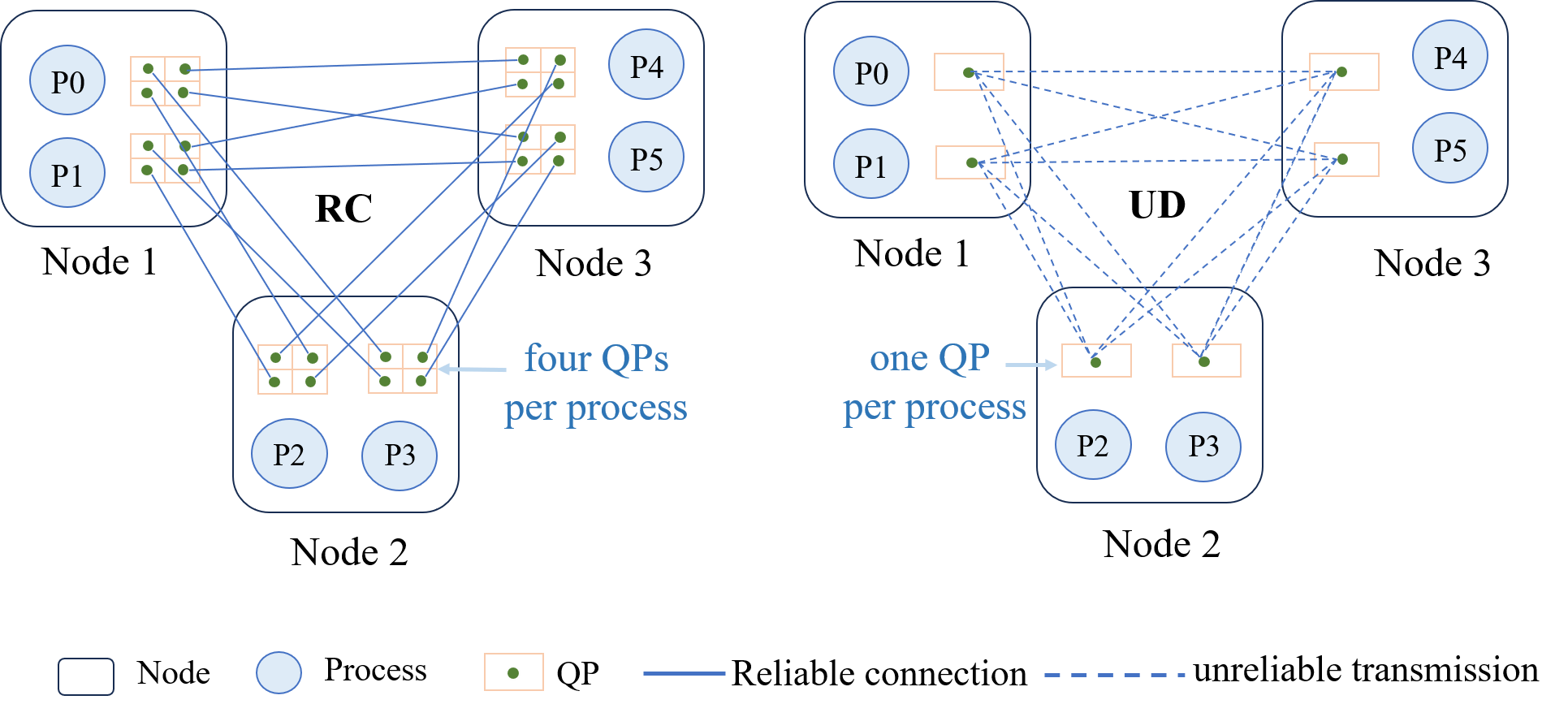}
\caption{QP growth in RC (left) vs. UD (right).}
\label{fig:qp-comparison}
\end{figure}
\label{sec:transport}
RDMA supports four QP  types: RC, Unreliable Connection (UC), UD and Reliable Datagram (RD)  are summarized in Table~\ref{tab:qp-comparison}.

\begin{table}[h]
\centering
\caption{Comparison of RDMA QP Transport Types}
\label{tab:qp-comparison}
\begin{tabular}{@{}lccc@{}}
\toprule
\textbf{Type} & \textbf{Connection} & \textbf{Reliability} & \textbf{Data Granularity} \\
\midrule
RC & Yes & Yes & Multi-packet supported \\
UC & Yes & No  & Multi-packet supported \\
UD & No  & No  & One MTU per WR \\
RD & No  & Yes & Undefined \\
\bottomrule
\end{tabular}
\end{table}
\begin{figure*}[t]
  \centering
  \subfloat[RNIC throughput]{%
    \includegraphics[width=0.24\linewidth]{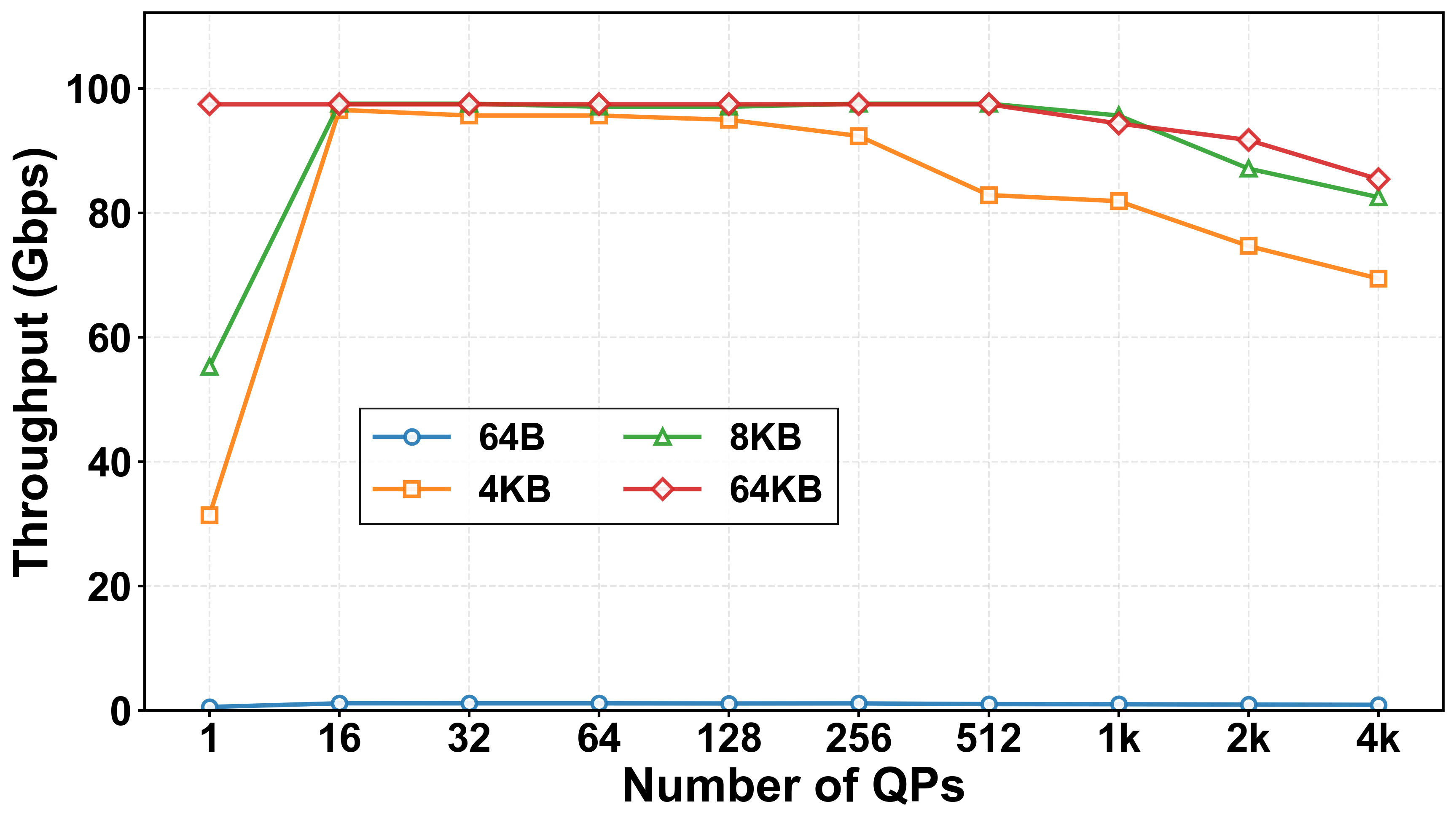}%
    \label{fig:ntarget-throughput}
  }
  \hfill
  \subfloat[Throughput degradation ratio]{%
    \includegraphics[width=0.24\linewidth]{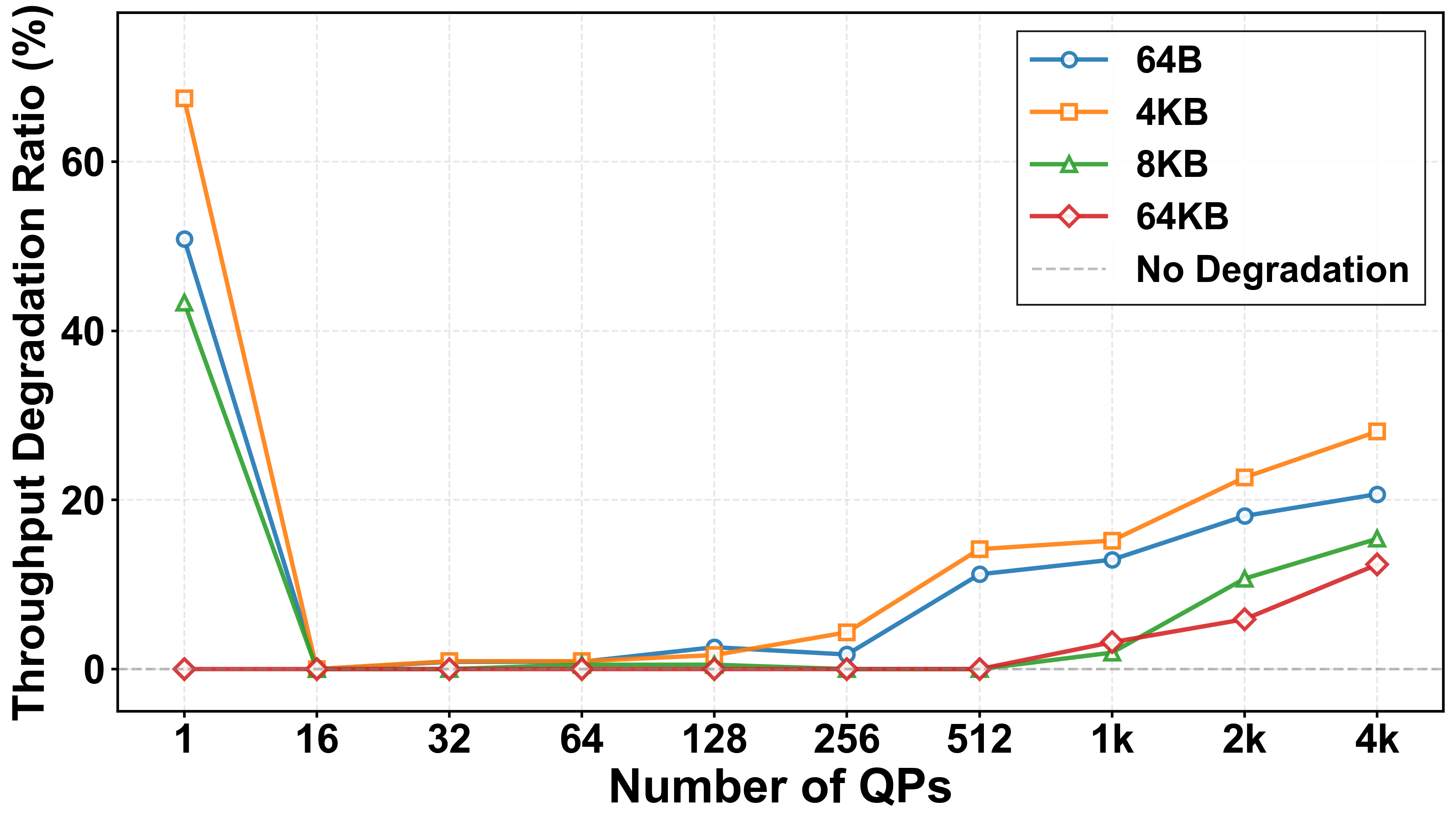}%
    \label{fig:ntarget-pcie}
  }
  \hfill
  \subfloat[Extra PCIe Bandwidth]{%
    \includegraphics[width=0.24\linewidth]{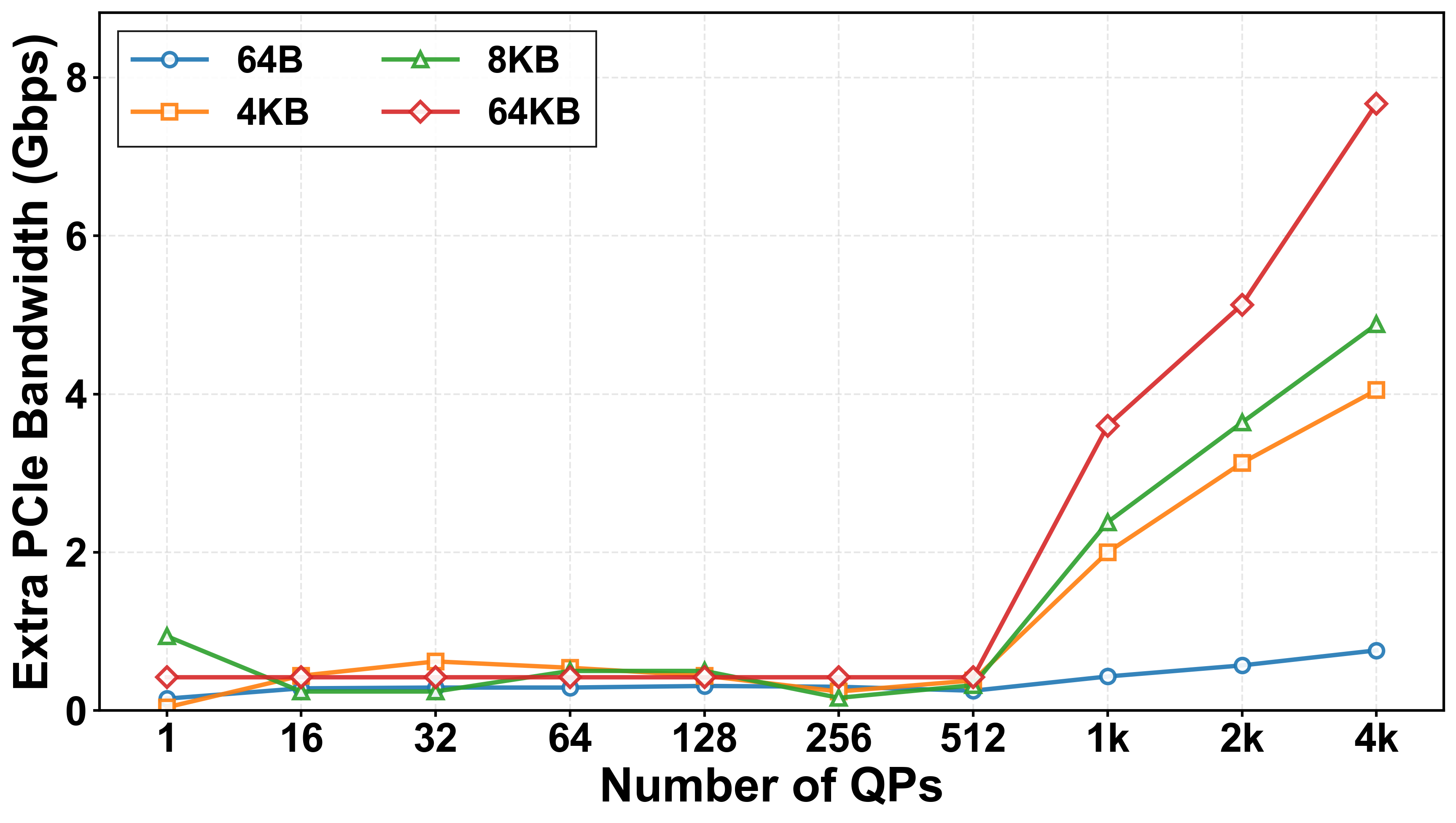}%
    \label{fig:txdepth}
  }
  \hfill
  \subfloat[Operations per second with different QPs]{%
    \includegraphics[width=0.24\linewidth]{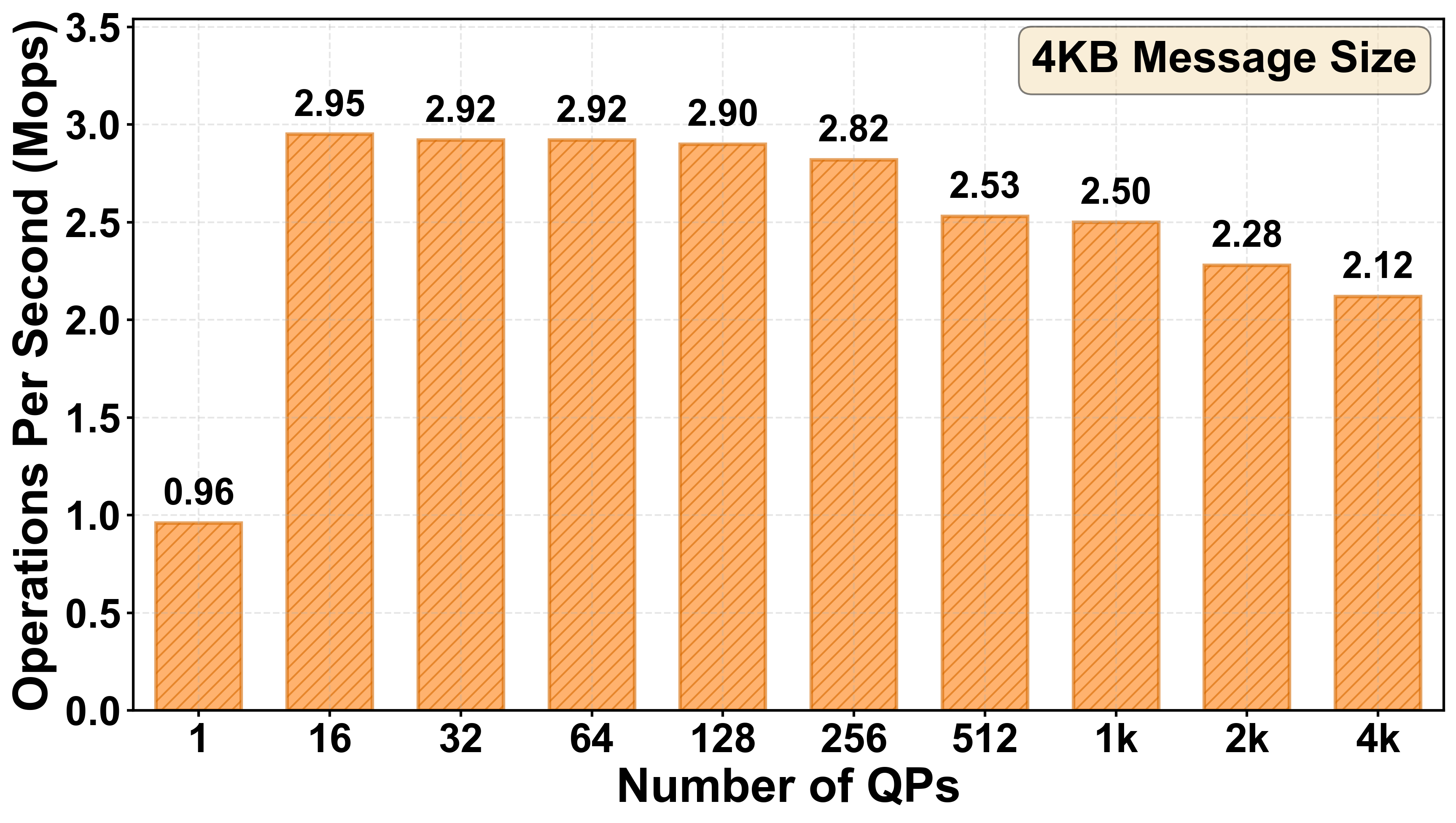}%
    \label{fig:schemes}
  }

  \caption{Scalability Issues of RNICs.}
  \label{fig:rnic-four}
\end{figure*}
RC is  widely deployed  for AI workloads requiring reliable delivery. However, it suffers from severe  scalability limitations. The left part of Fig.~\ref{fig:qp-comparison} illustrates the QP assignment  of RC. Suppose a communication group consists of $N$ nodes, each hosting $P$ processes. In RC mode, to allow all-to-all communication, each process must create $(N-1) \times P$ QPs to reach remote processes. Since all $P$ processes within a node share the same RNIC, the total number of QPs handled by a single RNIC grows quadratically as $(N-1) \times P^2$. As the number of nodes increases, the on-chip cache of the RNIC cannot accommodate all active QPs, which  leads to frequent QP cache misses.

 In contrast, as shown in the right of Fig~\ref{fig:qp-comparison}, UD mode decouples packet transmission  from connection state. Each process only needs one QP to send packets to any peer, keeping the total number of QPs per RNIC fixed at $P$, regardless of the cluster scale. While UD lacks reliability, it provides a foundation for scalable RDMA, so we base our CSA-UD atop the UD mode and an effective software-based reliability layer.

\subsection{RDMA Connection Scalability Issue}
\label{sec:scalability}
RDMA connection scalability issues are particularly pronounced in workloads with dense all-to-all communication patterns. A typical application scenario is Large Language Model (LLM) training, especially Mixture-of-Experts (MoE) training, where the communication process often involves  more than 10\,K\,QPs
. Such a large number of QPs pose significant challenges to RDMA connection management.
When a large number of QPs are created, constrained on-chip memory capacity prevents RNICs from caching all required QP metadata. Upon a cache miss, the RNIC must reload metadata from host memory via PCIe using DMA, introducing additional access latency and consuming valuable PCIe bandwidth, which in turn  degrades throughput. 

In order to quantitatively demonstrate the RDMA connection scalability issue, we conduct testbed  experiments using four hosts and one switch, where all hosts are equipped with Mellanox CX5 RNICs. The experimental topology consists of one server and three clients. All hosts are interconnected through a 100\,Gbps switch with 100\,Gbps links.
During the experiment, the server concurrently issues READ requests to all three clients over multiple RC connections to retrieve data blocks. To evaluate the scalability of the RNIC, we vary the number of concurrent RC connections and measure the aggregate throughput at the server, defined as the product of the  number of completed READ operations per second. In addition, since the server runs on an AMD CPU platform, we use AMD uProf to quantify the extra PCIe Bandwidth, which reflects the overhead caused by RNIC metadata cache misses.\footnote{
$\text{Extra PCIe Bandwidth} = \text{PCIe Throughput} - \text{RNIC Throughput}$.
}
The experimental results are presented in Fig.~\ref{fig:rnic-four}.

Fig.~\ref{fig:rnic-four} illustrates the scalability limitations of RNICs as the number of QPs increases, consistently observed across all message sizes.
As shown in Fig.~\ref{fig:rnic-four}(a), although RNIC throughput can approach the link rate at moderate QP counts(16--256), it inevitably degrades once the number of QPs exceeds 512. Specifically, throughput degradation manifests differently across message sizes. For 64\,B messages, throughput is fundamentally constrained by a low data-to-overhead ratio: per-packet processing and PCIe transaction costs remain nearly constant, leading to severe link underutilization and increased sensitivity to QP state thrashing. For 4\,KB messages operating near the MTU, throughput degradation is mainly attributed to limited pipeline depth, which cannot fully overlap metadata access latency when frequent QP context evictions occur. For 8\,KB and 64\,KB messages, deeper pipelines help amortize per-packet overhead and partially mask latency; however, throughput degradation still arises once the active QP working set exceeds the RNIC’s on-chip SRAM capacity, triggering excessive PCIe state fetches.
Fig.~\ref{fig:rnic-four}(b) further quantifies this trend, showing that throughput degradation ratio increases rapidly beyond a certain number of QPs for all tested  message sizes, indicating a fundamental scalability bottleneck rather than a message-size-specific effect.
Fig.~\ref{fig:rnic-four}(c) reveals the root cause of this degradation: the Extra PCIe Bandwidth increases sharply with the QP count across all message sizes, implying frequent metadata cache misses that force RNICs to fetch connection states from host memory via PCIe.
As a result, application-level efficiency is directly affected.
As shown in Fig.~\ref{fig:rnic-four}(d), even with a fixed 4\,KB message size, the achieved operation rate peaks at moderate QP counts and then steadily declines as the number of QPs further increases.
Overall, these results demonstrate that RDMA scalability issues are pervasive across message sizes and become pronounced under large-scale parallel communication.

RoUD~\cite{roud-ccgrid23}  and ERD~\cite{wang2024optimized} represent two attempts to build reliable RDMA transport over the UD mode. However, both of them exhibit critical limitations when applied to hyperscale AI training. The traditional Selective Repeat (SR) mechanism in RoUD fails to scale for large-scale collective communications, since it requires one ACK per packet, overwhelming the CQ and delaying loss detection. Besides, the bitmap-based NACK recovery in ERD mitigates some limitations of RoUD, but introduces new challenges such as spurious retransmission and kernel overhead introduced by TCP. The above issues and limitations motivate the design of CSA-UD, which incorporates adaptive loss detection, software-based loss recovery, and multipath-friendly lightweight reordering tailored for collective communication workload.
\section{Basic Idea}
\begin{figure}[t]
    \centering
    \includegraphics[width=0.9\linewidth]{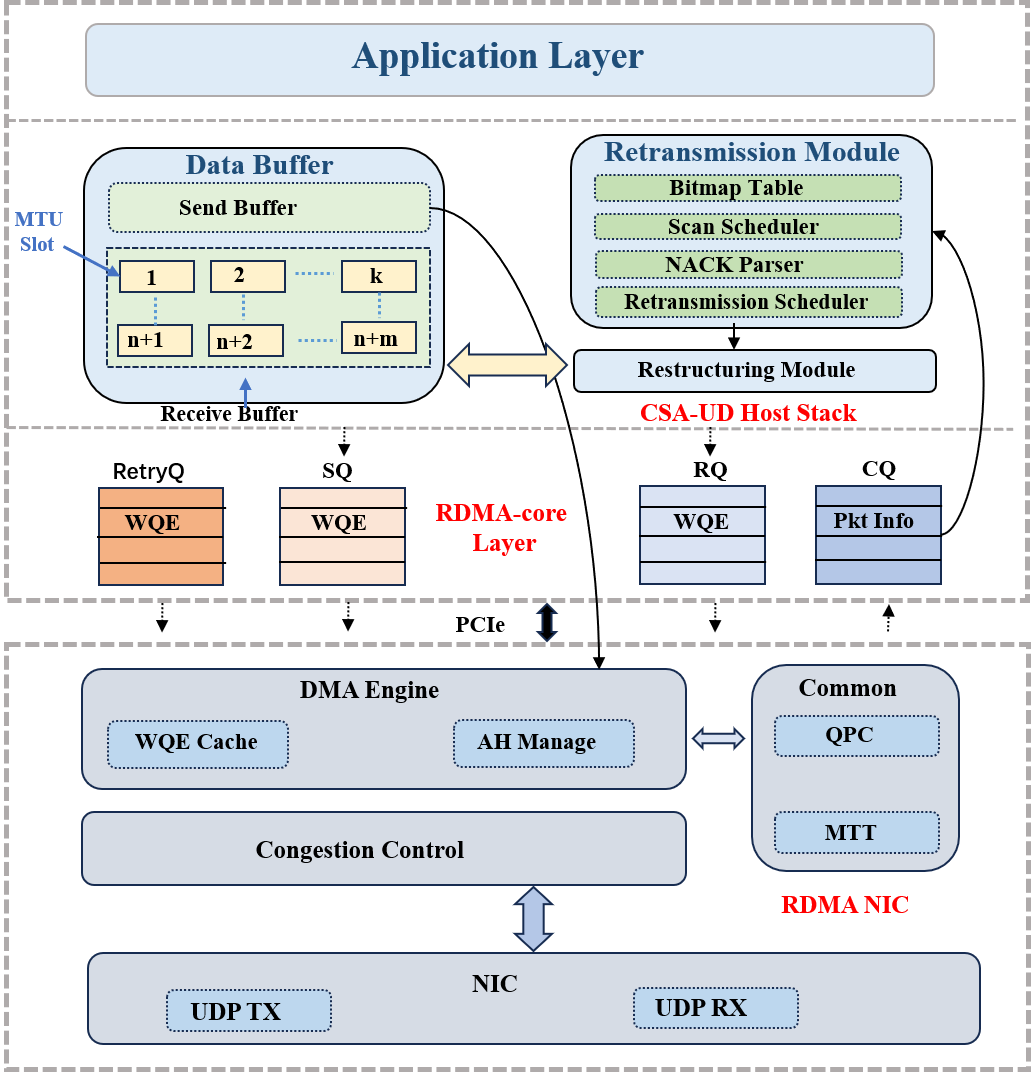}
    \caption{CSA-UD Framework.}
    \label{fig:framework}
\end{figure}
To address the limitations of existing RDMA reliability  mechanisms, i.e., QP scalability issue, tail loss recovery inefficiency and reordering overhead, CSA-UD proposes the following two key innovations.

\begin{enumerate}
\item \textbf{Tail-aware NACK via Adaptive Scanning.}
     CSA-UD introduces an adaptive scanning algorithm  for NACK-based loss detection. Rather than using a static scan interval, CSA-UD dynamically adjusts the scanning frequency based on two runtime indicators: remaining data volume, and  instantaneous receive rate. During the early transmission stages, the interval remains long to reduce control traffic and avoid spurious  losses due to out-of-order packets. In later stages, the scan interval shortens to quickly detect and retransmit tail packet losses. This strategy can reduce unnecessary signaling and  lowers tail latency.

\item \textbf{Bitmap-Guided One-Copy Reordering.}
To tolerate multipath-induced out-of-order packets  without excessive CPU and memory overhead, the receiver  leverages a lightweight reordering mechanism inspired by virtual memory systems. The receiver buffer is segmented into fixed-size, MTU-aligned slots. Packets are sequentially placed into available slots upon arrival, and a bitmap tracks the occupancy status of each slot~\cite{lu2017memory}. When the application requests data, CSA-UD scans the bitmap and assembles the complete message with a single memory copy. This eliminates the need for explicit packet sorting or multiple intermediate bufferings, preserving RDMA’s zero-copy semantics while enabling high-throughput, out-of-order reception.
\end{enumerate}

\section{DESIGN}
\begin{figure}[t]
  \centering
  \includegraphics[width=\linewidth]{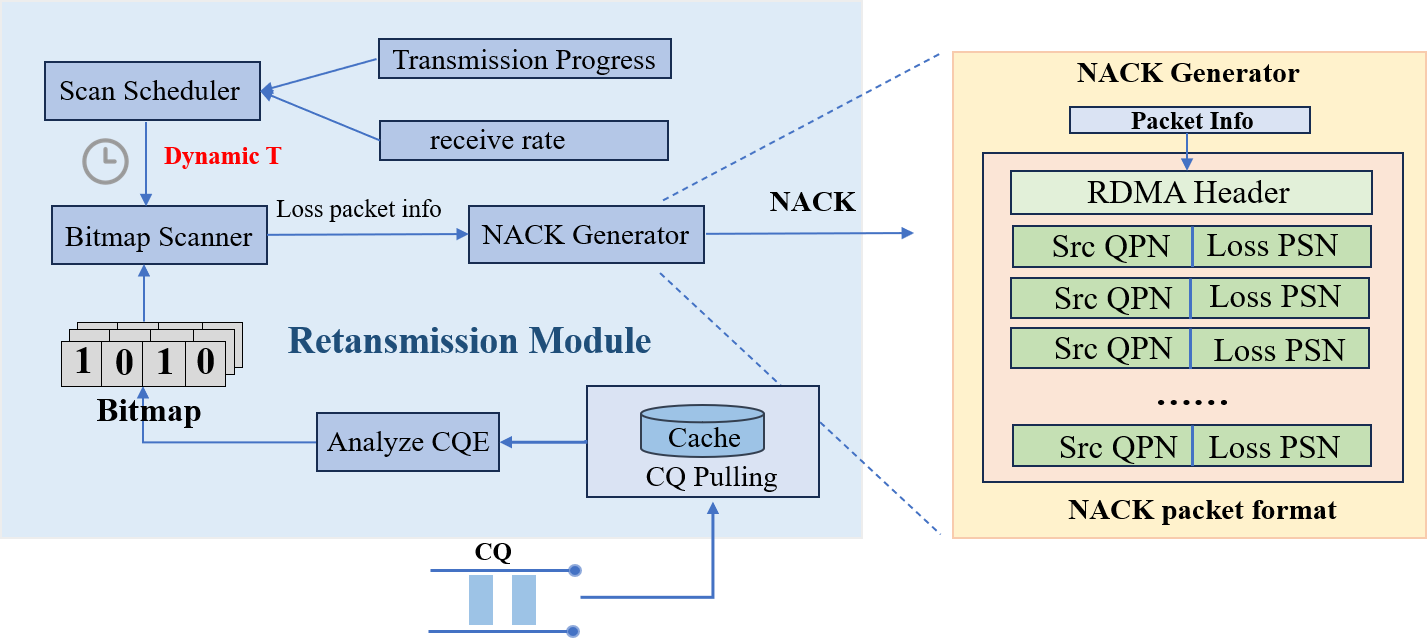}
  \caption{
    CSA-UD’s tail-aware NACK. 
    The scan scheduler adjusts the bitmap scan interval dynamically. 
    Lost packets are detected  and reported via selective NACKs to the sender.
  }
  \label{fig:nack-arch}
\end{figure}
\subsection{CSA-UD Overview }
 Besides QP scalability, CSA-UD has three additional design goals: 1) tail latency reduction to accelerate each training iteration, 2) efficient out-of-order packet reassembly to support multipath transmission, and 3) practical deployability on commodity RNICs without any reliance on PFC, reordering buffers, or custom hardware.

The CSA-UD framework is illustrated in Fig.~\ref{fig:framework}. All CSA-UD logic, including loss detection, recovery and packet reordering, is implemented entirely in software on the host, while the RNIC is only responsible for standard UDP send and receive operations. This design does not require any modification to the RNIC, thus has high deployability. 

To maximize reliability and reduce tail latency, CSA-UD assigns each training process two separate UD QPs: one data QP for transmitting original data packets and the other RetryQP dedicated to retransmitted packets and NACK feedback. During transmission, CSA-UD always prioritizes the RetryQP; only when it becomes empty are packets from the data QP sent. This design ensures loss recovery traffic is prioritized without interfering with regular transmission, thereby accelerating tail flow completion. After receiving a NACK, the sender immediately retransmits the lost packets via the RetryQP.

All control and data paths are directly managed via standard RDMA verbs (e.g., ibv\_post\_send, ibv\_post\_recv), without relying on any vendor-specific firmware extensions or software-defined layers. CSA-UD remains fully compatible with commodity RNICs that support UD mode.

\subsection{Tail-Aware NACK}
\label{sec:tail-aware-nack}

CSA-UD incorporates a receiver-side retransmission module to detect packet losses and trigger recovery via NACKs. Specifically, CSA-UD implements an  adaptive loss scan interval mechanism as well as the packet retransmission logic  in userspace that are  decoupled from the normal data path, minimizing PCIe overhead and avoiding NACK storms.

As illustrated in Fig.~\ref{fig:nack-arch}, CSA-UD employs a dedicated receiver thread that continuously polls Completion Queue Elements (CQEs) from the RDMA-core layer and updates a bitmap  indexed by (SrcIP, SrcQPN, DstQPN, PSN), where each bit indicates the reception status of a specific packet; successfully received packets are marked as 1. Besides, a  software timer, whose period is dynamically adjusted according to transmission progress and receive rate, periodically scans the bitmap to identify missing packets and trigger retransmissions through NACKs.

To reduce traffic through PCIe  and CPU load, CSA-UD avoids generating an ACK per packet. Instead, only missing packets are selectively reported, each carrying the (QPN, PSN) tuple, where QPN denotes the  queue pair number and PSN represents the packet sequence number.

\begin{figure}[t]
  \centering
  \includegraphics[width=\linewidth]{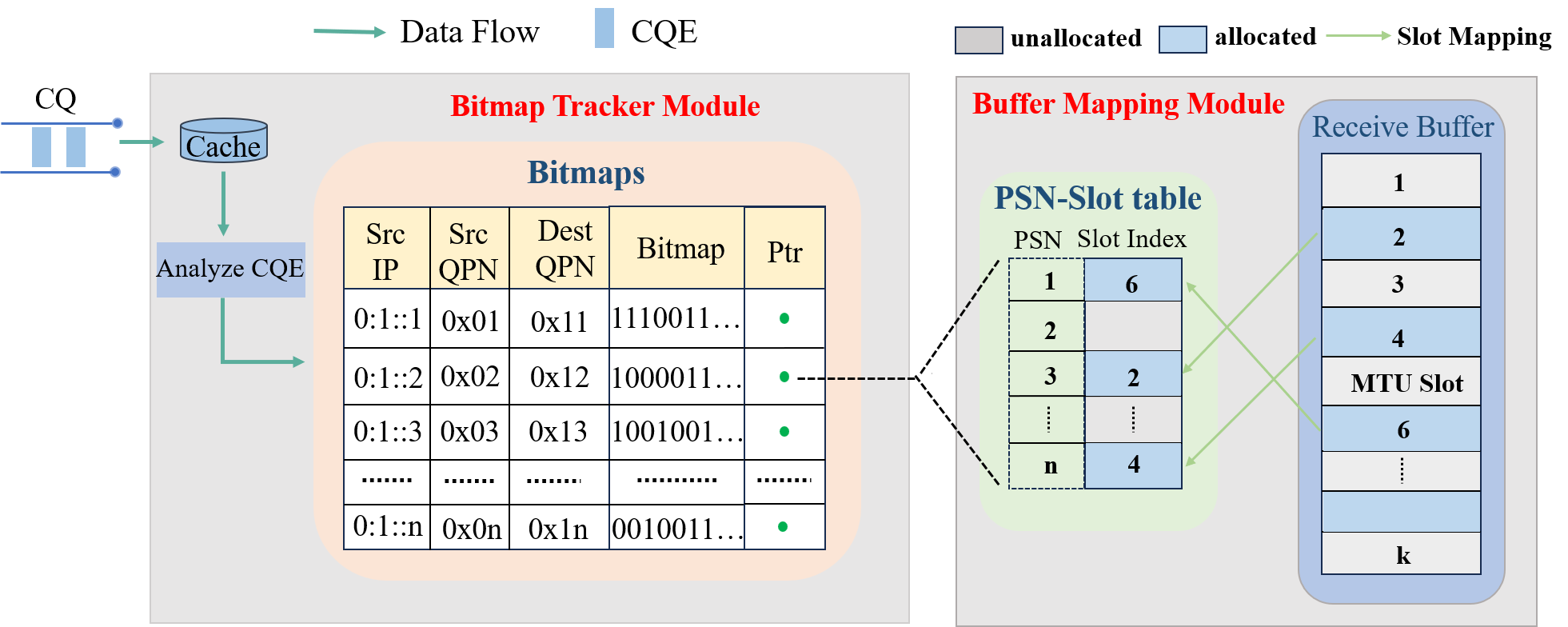}
  \caption{
    Receiver-side bitmap and buffer mapping logic. 
    PSNs are mapped to fixed buffer slots for zero-copy placement.
  }
  \label{fig:buffer-mapping}
\end{figure}

As shown in Fig.~\ref{fig:buffer-mapping}, upon polling a CQE, the receiver extracts (SrcIP, SrcQPN, PSN) and locates the corresponding entry in the bitmap. Each flow is uniquely identified by (SrcIP, SrcQPN, DstQPN), and the bitmap marks received PSNs. The CQ polling unit is equipped with a local cache to prevent CQ overflow when packet arrivals exceed the  processing rate.

When a packet arrives, its PSN is marked, and a dedicated pointer in the bitmap  updates a precomputed PSN-to-slot mapping table. This table associates each PSN with a preallocated receive buffer slot, enabling direct placement of incoming packets into their designated slots, even under out-of-order delivery conditions. This design avoids intermediate buffering or reordering queues, thereby reducing memory copy overhead. The final in-order reassembly is deferred to a dedicated reassembly engine, effectively decoupling CQ polling, loss detection, and buffer management operations.

In collective communication patterns, the total data volume expected per iteration, denoted as $D_{\text{total}}$, is generally known a priori due to the synchronized nature of collective communication. Specifically, in each communication round $i$, each node receives data from a subset of peer processes, whose number is denoted as $P_i$. For the receiving node, $P_i$ can be dynamically obtained from the number of valid rows in the bitmap it maintains, which reflects the set of peers actively communicating with it during round $i$. Let $D_{\text{send}}$ represent the amount of data each peer is expected to send to this node. Then the total expected received data volume in round $i$ is:
\begin{equation}
    D_{\text{total}} = P_i \cdot D_{\text{send}}
\end{equation}

Let $N_{\text{recv},i}$ represent the cumulative number of packets received by scan round $i$. The residual data volume at round $i$ can then be expressed as:
\begin{equation}
    D_{\text{rem},i} = D_{\text{total}} - N_{\text{recv},i} \cdot S_{\text{payload}}
\end{equation}
where $S_{\text{payload}}$ denotes the effective payload size per packet, computed as:
\begin{equation}
    S_{\text{payload}} = \text{MTU} - H_{\text{proto}} - H_{\text{user}} - H_{\text{icrc}}
\end{equation}
Here, $H_{\text{proto}}$ represents the cumulative protocol header size, including fixed components such as IP, UDP, BTH, and DETH headers, with a total size of 48\,B for IPv4 stacks and 68\,B for IPv6 stacks. $H_{\text{user}}$ represents user-defined metadata overhead and $H_{\text{icrc}}$ is a 4\,b ICRC checksum field.

\subsection{Adaptive Bitmap Scan Interval}

The bitmap scan interval $T_i$ governs the trade-off between loss detection latency and CPU overhead. A shorter interval enables faster NACK generation but increases polling frequency, while a longer interval conserves CPU cycles at the cost of delayed retransmissions. To achieve an optimal balance, CSA-UD dynamically adjusts $T_i$ by integrating transmission progress awareness with network condition feedback.

\subsubsection{Progress-Aware Base Interval}

Define the transmission progress ratio at round $i$ as:
\begin{equation}
    \rho_i = 1 - \frac{D_{\text{rem},i}}{D_{\text{total}}} = \frac{N_{\text{recv},i} \cdot S_{\text{payload}}}{D_{\text{total}}} \in [0, 1]
\end{equation}
where $\rho_i$ monotonically increases from 0 to 1 as transmission progresses, and all other symbols have been defined in Equations~(1), (2), and (3).

The key insight motivating progress-aware adaptive interval adjustment is the \emph{asymmetric urgency} of loss detection across transmission phases. During early transmission ($\rho_i \approx 0$), undetected packet losses can be recovered in subsequent scan rounds with minimal impact on completion time, as the majority of data transfer remains ahead. Such practice also helps mitigate spurious loss detection caused by packet reordering. However, as transmission approaches completion ($\rho_i \approx 1$), any remaining losses become critical path items---they directly determine the final flow completion time and may block dependent collective operations. This asymmetry motivates progressively more aggressive scanning as flow transmission advances.

In this context, define the progress-aware interval excess as:
\begin{equation}
    \Delta T_i = (T_{\max} - T_{\min}) \cdot (1 - \rho_i^\alpha)
\end{equation}
where $T_{\min}$ and $T_{\max}$ are the scan interval bounds, and $\alpha > 0$ controls the decay profile. The term $\Delta T_i$ represents the additional interval beyond $T_{\min}$: it equals $T_{\max} - T_{\min}$ when $\rho_i = 0$ and vanishes as $\rho_i \to 1$, ensuring progressively aggressive scanning near completion.

\subsubsection{Rate-Adaptive Modulation}

Progress-aware scheduling captures the temporal dynamics of flow transmission but remains agnostic to instantaneous network conditions. Under congestion or bursty packet losses, the progress-based interval alone may be suboptimal. To address this, we introduce a rate-based modulation factor $F_i$.

The rate modulation factor quantifies the deviation between observed and expected receiving rates:
\begin{equation}
    F_i = \left( \frac{R_{\text{ideal}}}{{R}_i + \epsilon} \right)^\gamma
\end{equation}
where $R_{\text{ideal}}$ denotes the ideal receive rate, which is set to the link capacity and is known a priori from the hardware specification. The parameter $\epsilon$ is a small constant for numerical stability. The exponent $\gamma \in (0,1]$ controls the sensitivity of the scan interval to throughput deviation from the ideal rate. It is derived from two operational parameters: the maximum expected degradation ratio $\kappa = R_{\text{ideal}}/R_{\min}$ and the maximum allowed modulation factor $F_{\max}$. Enforcing $F_i \leq F_{\max}$ at $R_i = R_{\min}$ gives
\begin{equation}
\left(\frac{R_{\text{ideal}}}{R_{\min}}\right)^\gamma = F_{\max}
\quad \Rightarrow \quad
\gamma = \frac{\ln F_{\max}}{\ln \kappa}.
\end{equation}
For example, in a 100\,Gbps network with $R_{\min}=20$\,Gbps ($\kappa=5$), choosing $F_{\max}=\sqrt{5}$ yields $\gamma=0.5$. Smaller $\gamma$ results in a sub-linear, dampened response, while $\gamma=1$ produces a linear modulation. When the receiving rate falls below expectation (${R}_i < R_{\text{ideal}}$), we have $F_i > 1$, which extends the scan interval to reduce CPU overhead; high throughput (${R}_i \geq R_{\text{ideal}}$) yields $F_i \leq 1$, enabling more frequent scanning to promptly detect and recover losses.

\subsubsection{Composite Interval}

The final scan interval $T_i$ integrates progress awareness with rate-adaptive modulation:
\begin{equation}
    \hat{T}_i = T_{\min} + \Delta T_i \cdot F_i
\end{equation}
The modulation factor $F_i$ scales only the excess interval $\Delta T_i$, not the minimum bound $T_{\min}$. This design guarantees convergence: as $\rho_i \to 1$, $\Delta T_i \to 0$ regardless of $F_i$, ensuring $\hat{T}_i \to T_{\min}$. Consequently, the system always reaches the maximum scanning frequency near flow completion, regardless of network conditions.

The final interval is bounded within $[T_{\min}, T_{\max}]$:
\begin{equation}
T_i = \operatorname{clamp}(\hat{T}_i,\; T_{\min},\; T_{\max}) = 
\begin{cases}
T_{\min}, & \text{if } \hat{T}_i < T_{\min} \\
T_{\max}, & \text{if } \hat{T}_i > T_{\max} \\
\hat{T}_i, & \text{otherwise}
\end{cases}
\end{equation}
The upper bound prevents excessive idling under severe degradation, while the lower bound avoids wasteful over-polling. Together with the convergence property, this ensures that $T_i$ transitions smoothly from conservative scanning at the beginning of transmission to aggressive scanning at the end, with rate-adaptive adjustments modulating the transition speed according to real-time network conditions.

\begin{figure}[t]
  \centering
  \includegraphics[width=0.85\linewidth]{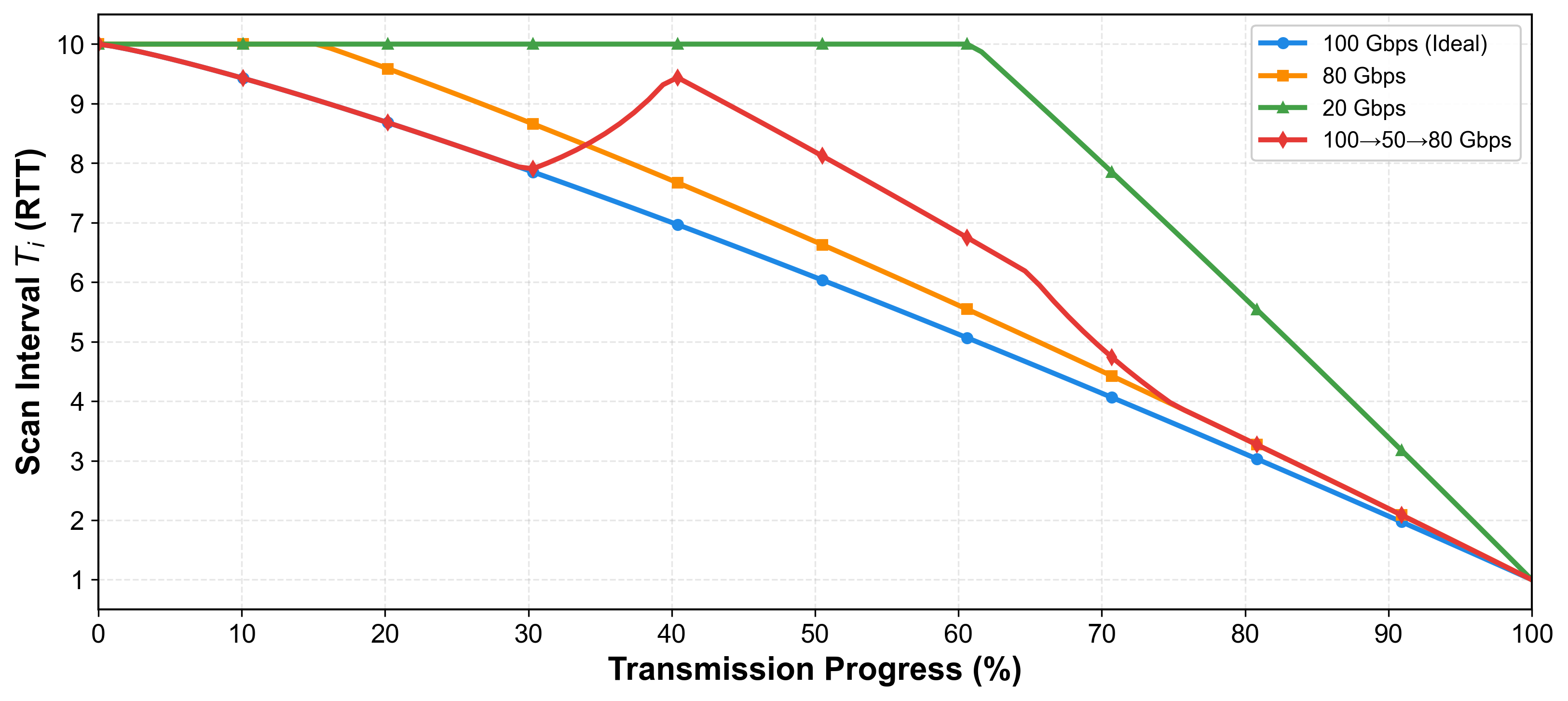}
  \caption{
    Scan interval $T_i$ adaptation under varying bandwidth conditions.
  }
  \label{fig:scan-interval-curves}
\end{figure}

As shown in Fig.~\ref{fig:scan-interval-curves}, CSA-UD reduces $ T_i $ as transmission progresses. Under 100\,Gbps (solid blue curve), the scan interval drops from 10\,RTT to 1\,RTT; under lower bandwidth conditions (80\,Gbps and 20\,Gbps), the interval decreases more conservatively to reduce overhead, which includes CPU load from frequent bitmap scans, PCIe traffic for control messages, and potential redundant retransmissions caused by premature loss detection. The solid red line with diamond markers demonstrates adaptive behavior: In the dynamic scenario (100$\to$50$\to$80\,Gbps), bandwidth drops linearly from 100\,Gbps to 50\,Gbps over 30\%--40\% progress and stays at 50\,Gbps until 65\%. The scan interval therefore increases from 30\% as $F_i$ grows with the deviation from $R_{\text{ideal}}$. From 65\% to 75\%, bandwidth recovers to 80\,Gbps, reducing $F_i$ and the interval accordingly. Beyond 75\%, bandwidth stabilizes at 80\,Gbps, and the interval converges toward $T_{\min}$ under the dominant progress factor. We set  $ T_{\max} = $ 10\,RTT  to avoid premature detection of delayed arriving packets due to congestion or out-of-order packets due to multipath transmission, which could cause unnecessary spurious  retransmissions. These spurious retransmissions not only waste network bandwidth but also introduce additional network load, further exacerbating congestion.  The lower bound is  $T_{\min}$ is set to 1\,RTT to ensure rapid detection of tail loss.

\subsection{Packet Format Design}
\begin{figure}[t]
  \centering
  \includegraphics[width=0.95\linewidth]{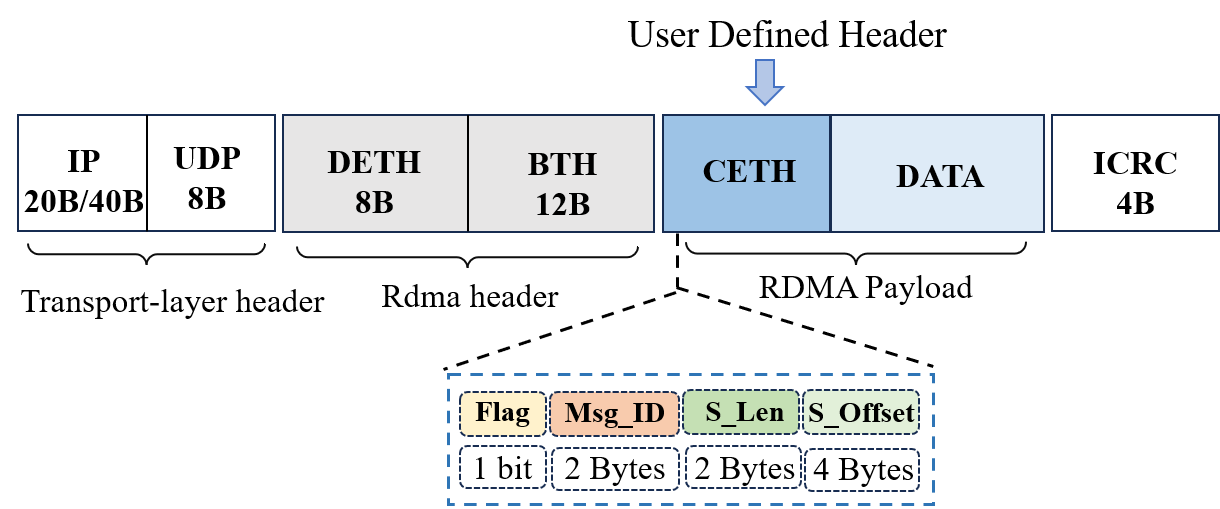}
  \caption{CSA-UD packet format with CETH}
  \label{fig:packet-structure-csaud}
\end{figure}

CSA-UD is built on the standard UD  mode defined by the IBTA specification~\cite{IBTA_Vol1_1_8}. To support host-side reliability and out-of-order reassembly, CSA-UD introduces a lightweight Custom Extension Header (CETH) inserted at the beginning of  RDMA payload. The full packet structure is shown in Fig.~\ref{fig:packet-structure-csaud}.

The packet header consists of the following fields: Transport layer header includes the IP header (20\,B for IPv4 or 40\,B for IPv6) and UDP header (8\,B), used for basic datagram routing. RDMA header includes: BTH (Base Transport Header, 12\,B), which encodes the QP number, PSN (Packet Sequence Number), and RDMA operation codes;
and DETH (Datagram Extended Transport Header, 8\,B), which is used only in UD mode. CETH (9\,B) is placed at the beginning of the RDMA payload and enables software-based reliability and reordering. It contains:
a 1bit flag indicating whether the packet is the last fragment of a message (set to 1 if no more segments follow for the same Msg\_ID),
a 2\,B Msg\_ID field identifying the associated message,
a 2\,B S\_Len field specifying the actual valid payload size,
and a 4\,B S\_Offset field indicating the byte offset within the original message.

\textbf{Overhead Analysis:} The 9\,B CETH imposes minimal overhead on  4\,KB MTU packets. It enables user-level reassembly and reliable transmission, eliminating the need for NIC-side reordering buffers and supporting efficient multipath transfer.

\subsection{ Bitmap-Guided One-Copy Reordering Mechanism.}
The UD transport service in RDMA inherently supports receiving out-of-order packets from multiple sources, a common scenario in multi-path environments. 
State-of-the-art reassembly mechanisms, such as ERD, address this by first buffering all incoming packets before performing reordering.
However, these approaches suffer from significant inefficiencies.
For instance, ERD processes packets in multiple steps: it first performs a pass over the buffered packets to classify them by source, followed by a second pass to group them by message. 
This process culminates in a final sorting step to reconstruct  message data before delivering it to the application.

This multi-pass methodology incurs substantial performance penalties.
It causes high I/O amplification due to repeated memory access for classification and data movement.
This not only leads to inefficient utilization of receive buffers but also exacerbates PCIe contention and introduces significant memory latency.
Even on resource-rich AI infrastructures, such latency can become a critical performance bottleneck.
\begin{algorithm}[t]
\caption{Per-Flow Reordering Algorithm}
\label{alg:csa-ud-reassembly}
\begin{algorithmic}[1]
\Require PSN-to-Slot table $\mathcal{T}$, packet buffer $\mathcal{B}$
\Ensure Ordered application data $\mathrm{DATA}$

\State Initialize $\mathrm{MsgLen} \gets \{\}$, $\mathrm{MsgAddr} \gets \{\}$, $\mathrm{Offset} \gets 0$

\Statex \Comment{\textit{Phase 1: Compute total length per message}}
\For{each $\mathrm{slot} \in \mathcal{T}$}
    \State $\mathrm{addr} \gets \textsc{GetAddr}(\mathrm{slot})$
    \State $\mathrm{hdr} \gets \textsc{ReadHeader}(\mathrm{addr})$
    \State $\mathrm{MsgLen}[\mathrm{hdr.MsgID}] \mathrel{+}= \mathrm{hdr.SegLen}$
\EndFor

\Statex \Comment{\textit{Phase 2: Compute destination offset per message}}
\For{each $\mathrm{msgID} \in \mathrm{MsgLen}$ in ascending order}
    \State $\mathrm{MsgAddr}[\mathrm{msgID}] \gets \mathrm{Offset}$
    \State $\mathrm{Offset} \gets \mathrm{Offset} + \mathrm{MsgLen}[\mathrm{msgID}]$
\EndFor

\Statex \Comment{\textit{Phase 3: Copy payload to application buffer}}
\For{each $\mathrm{slot} \in \mathcal{T}$}
    \State $\mathrm{addr} \gets \textsc{GetAddr}(\mathrm{slot})$
    \State $\mathrm{pkt} \gets \textsc{ReadPacket}(\mathrm{addr})$
    \State $\mathrm{hdr} \gets \textsc{ParseHeader}(\mathrm{pkt})$
    \State $\mathrm{dst} \gets \mathrm{AppBuf} + \mathrm{MsgAddr}[\mathrm{hdr.MsgID}] + \mathrm{hdr.SegOffset}$
    \State $\textsc{Write}(\mathrm{dst}, \mathrm{pkt.payload})$
\EndFor

\Statex \Comment{\textit{Phase 4: Verify completeness}}
\State $\mathrm{DATA} \gets \textsc{ReadBuffer}(\mathrm{AppBuf}, \mathrm{Offset})$
\If{$|\mathrm{DATA}| = \mathrm{Offset}$}
    \State \Return $\mathrm{DATA}$
\Else
    \State \Return \texttt{ERROR: incomplete}
\EndIf
\end{algorithmic}
\end{algorithm}

To overcome the above limitations, CSA-UD proposes a bitmap-guided one-copy packet reordering mechanism that enables highly efficient reassembly through a technique termed \textit{Direct Payload Placement}. Our approach is built on two key principles: leveraging the fixed-size nature of UD packets and employing a page-table-inspired mapping scheme for buffer management. This design eliminates the need for intermediate buffering and avoids costly multi-pass sorting.

To formalize the mechanism, Algorithm 1 outlines the complete reassembly process in CSA-UD. The reordering mechanism operates as follows. Upon a packet’s arrival, the NIC’s DMA engine places it directly into a pre-allocated receive buffer. Rather than immediately reading the full packet, the CSA-UD reassembly module first identifies its memory slot via the completion queue entry (CQE) and records it in a lightweight PSN-to-Slot mapping table.

Once transmission completes, CSA-UD initiates the reassembly procedure. 
For each flow, CSA-UD first locates its PSN-to-Slot mapping table using the 3-tuple identifier (Src\_IP, Src\_QPN, Dst\_QPN). 
Reassembly then proceeds by referencing this mapping table to retrieve all received packets of the flow.

In the first pass, CSA-UD performs a lightweight header-only memory access, reading only the 9-byte CETH to extract metadata including the message ID (Msg\_ID), segment offset (S\_Offset), and segment length (S\_Len). 
This header-level access introduces minimal memory traffic compared to full payload reads. 
The collected metadata is used to accumulate the total length of each message and to prepare for subsequent completeness verification (Algorithm~\ref{alg:csa-ud-reassembly}, lines~2--6).

In the second pass, CSA-UD computes the destination base address of each message in the application buffer using the accumulated message lengths (lines~7--9). 
It then reads the payload of each packet and calculates the  destination address by combining the message base address with the segment offset (lines~11--15). 
The payload is directly copied to its final location without explicit sorting or temporary staging buffers.

Overall, CSA-UD reduces reassembly to two sequential memory passes and a single memory write per packet. 
The first pass accesses only the compact CETH header, thereby imposing minimal pressure on memory bandwidth. 
This reassembly logic is encapsulated in a transparent module below the application layer, enabling seamless integration with existing AI frameworks while maintaining high performance.
\section{EVALUATION}
To evaluate the performance of CSA-UD, we first conduct testbed experiments on a real RDMA cluster to assess its basic throughput, then perform large-scale simulations in the ns-3 simulator to evaluate its flow-level performance.

\subsection{Testbed Experiments}

\textbf{Experimental Setup.}~
The experimental topology is identical to that described in Section~\ref{sec:scalability}, consisting of four hosts interconnected via a single 100\,Gbps switch with 100\,Gbps links. We conduct two separate sets of experiments: one using RC queue pairs and the other using UD queue pairs. In RC mode, the server concurrently issues RDMA READ requests to all three clients over multiple connections with a transmit depth of 32. In UD mode, communication is performed via send/receive operations, with the flow direction the same as in RC.

\begin{figure}[t]
  \centering
  \includegraphics[width=0.95\linewidth]{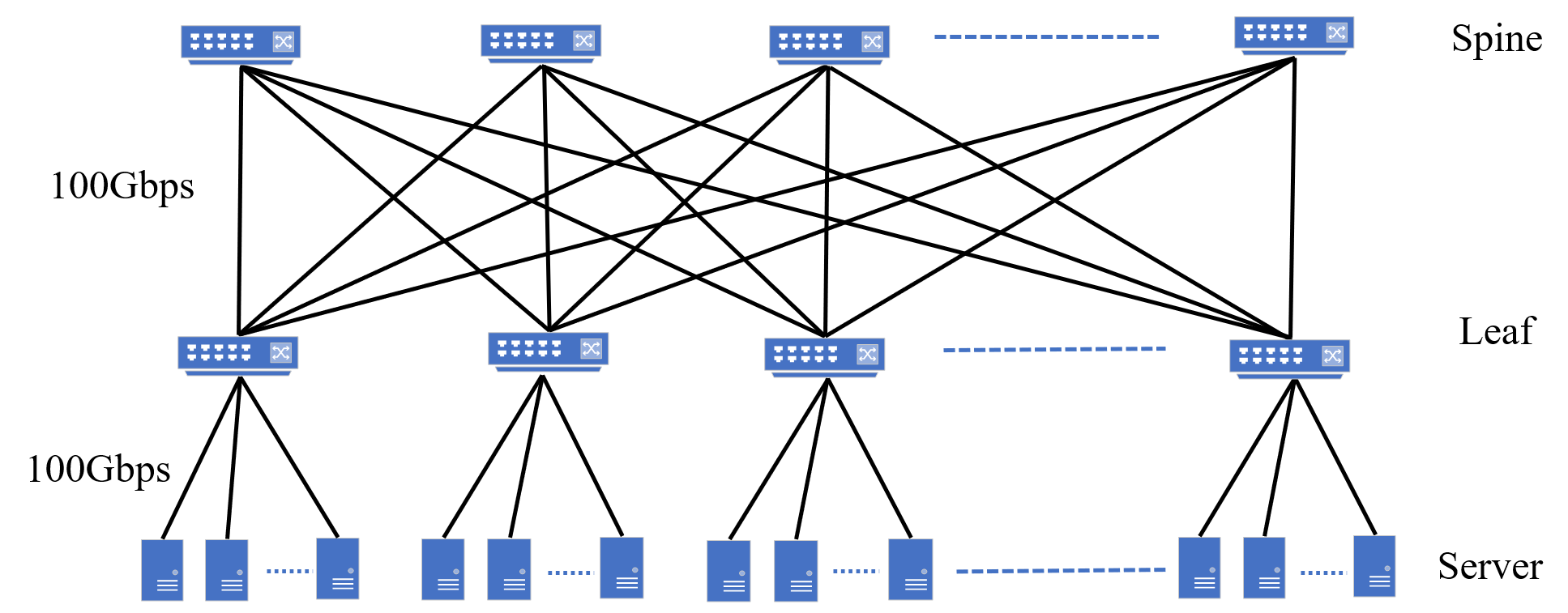}
  \caption{Leaf-spine topology used in simulations.}
  \label{fig:topo}
\end{figure}
\begin{figure*}[t]
  \centering
  \subfloat[QP Number=512]{%
    \includegraphics[width=0.24\linewidth]{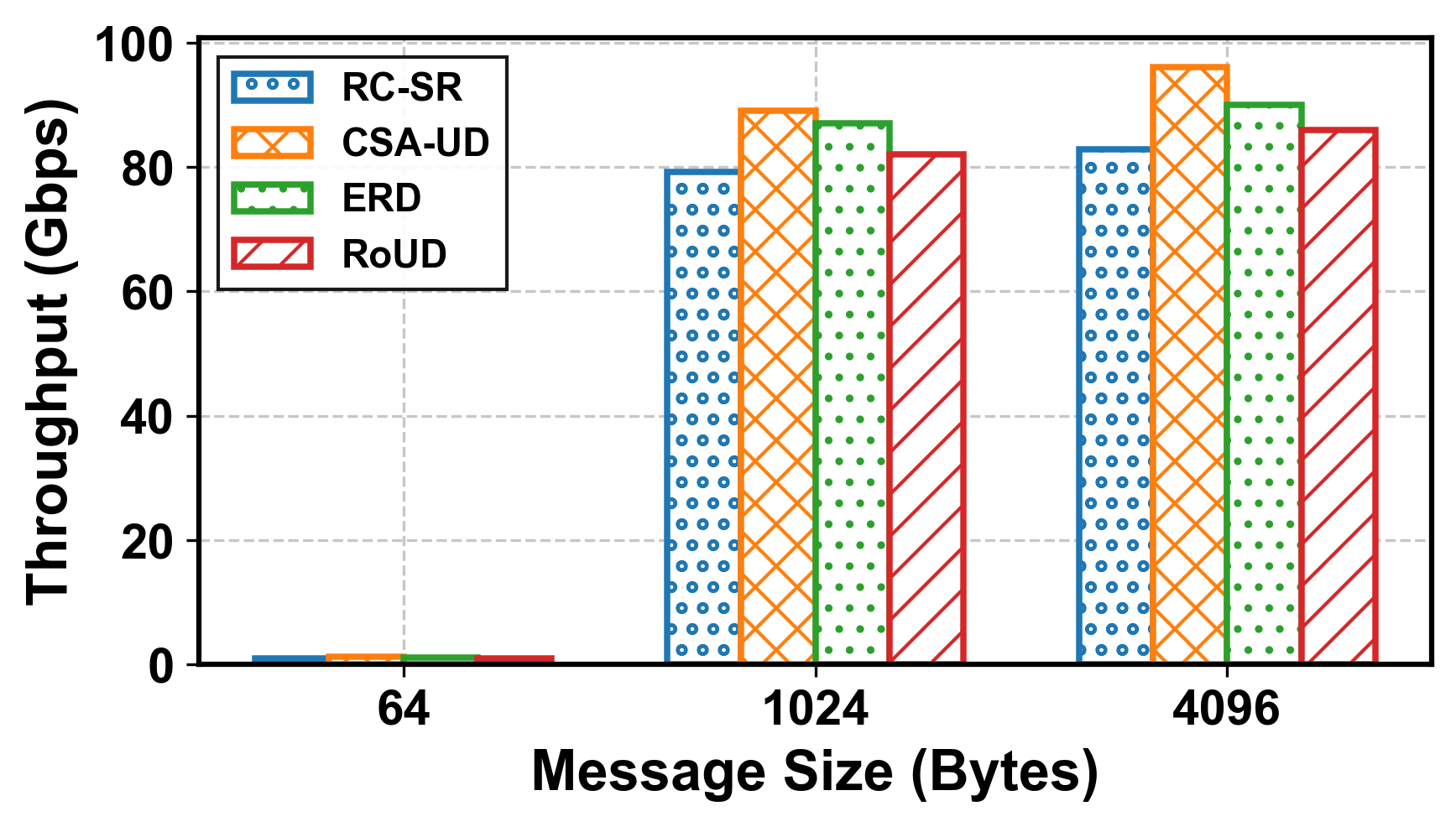}%
    \label{fig:ntarget-throughput}
  }
  \hfill
  \subfloat[QP Number=1024]{%
    \includegraphics[width=0.24\linewidth]{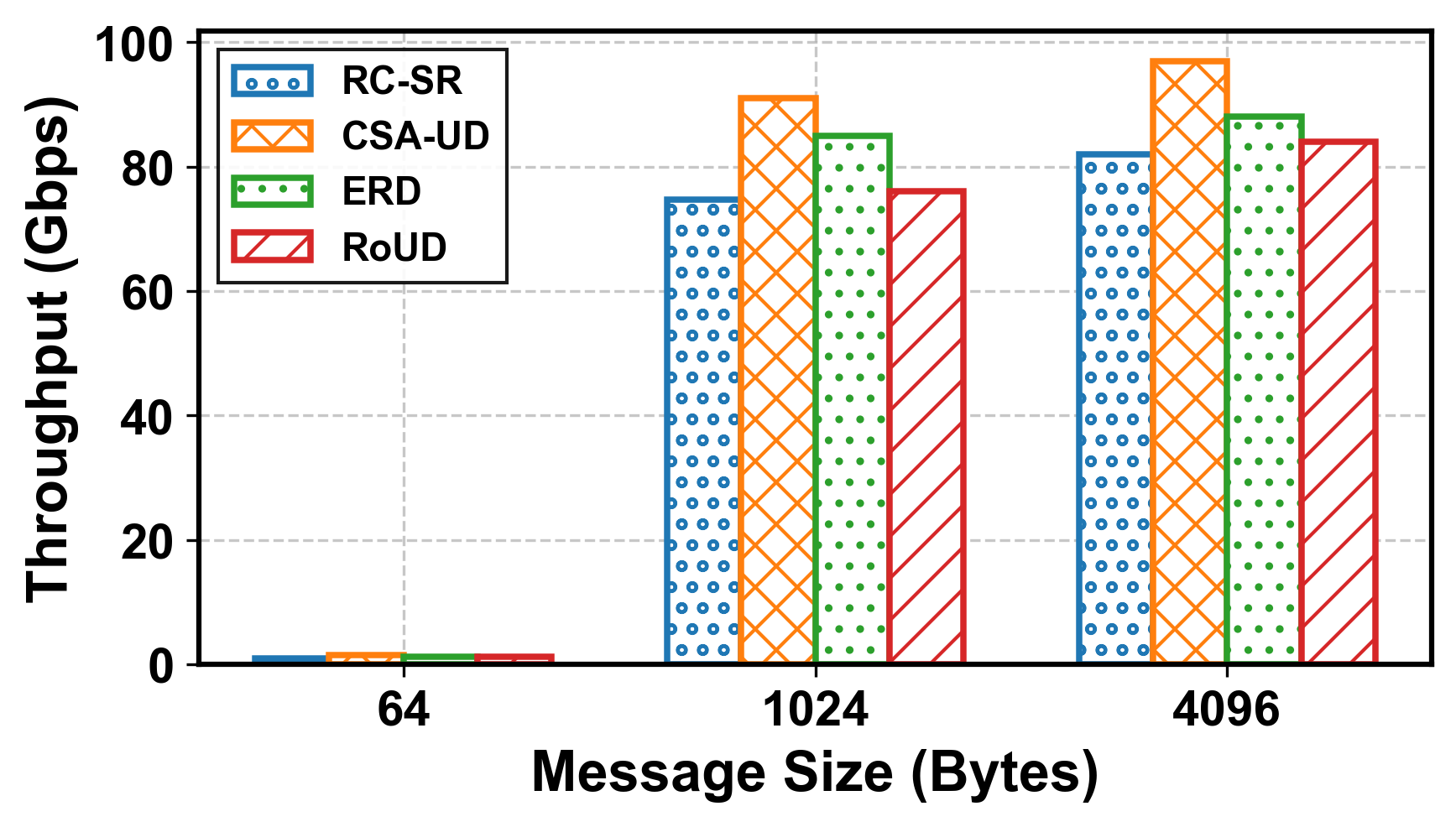}%
    \label{fig:ntarget-pcie}
  }
  \hfill
  \subfloat[QP Number=2048]{%
    \includegraphics[width=0.24\linewidth]{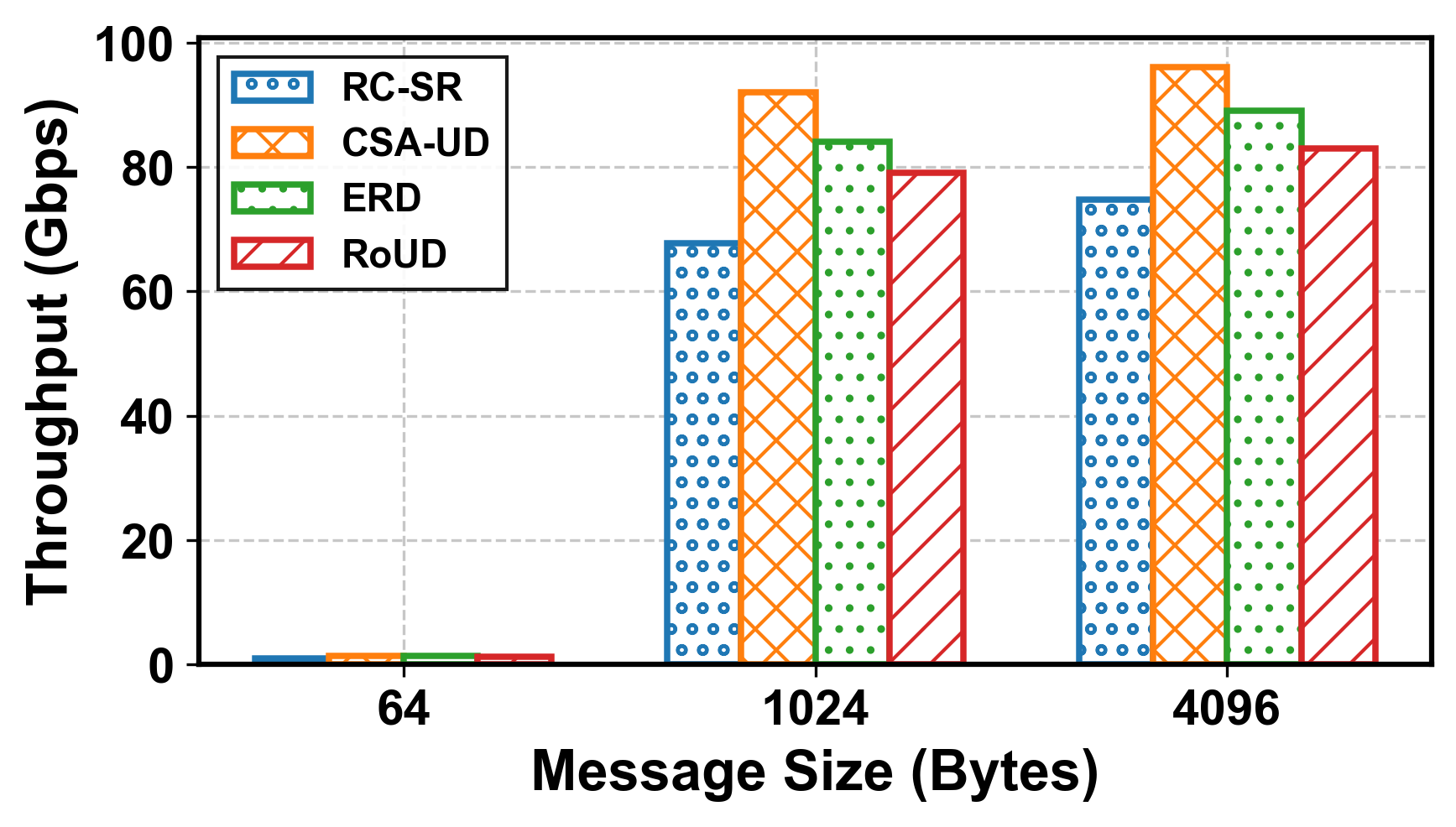}%
    \label{fig:txdepth}
  }
  \hfill
  \subfloat[QP Number=4096]{%
    \includegraphics[width=0.24\linewidth]{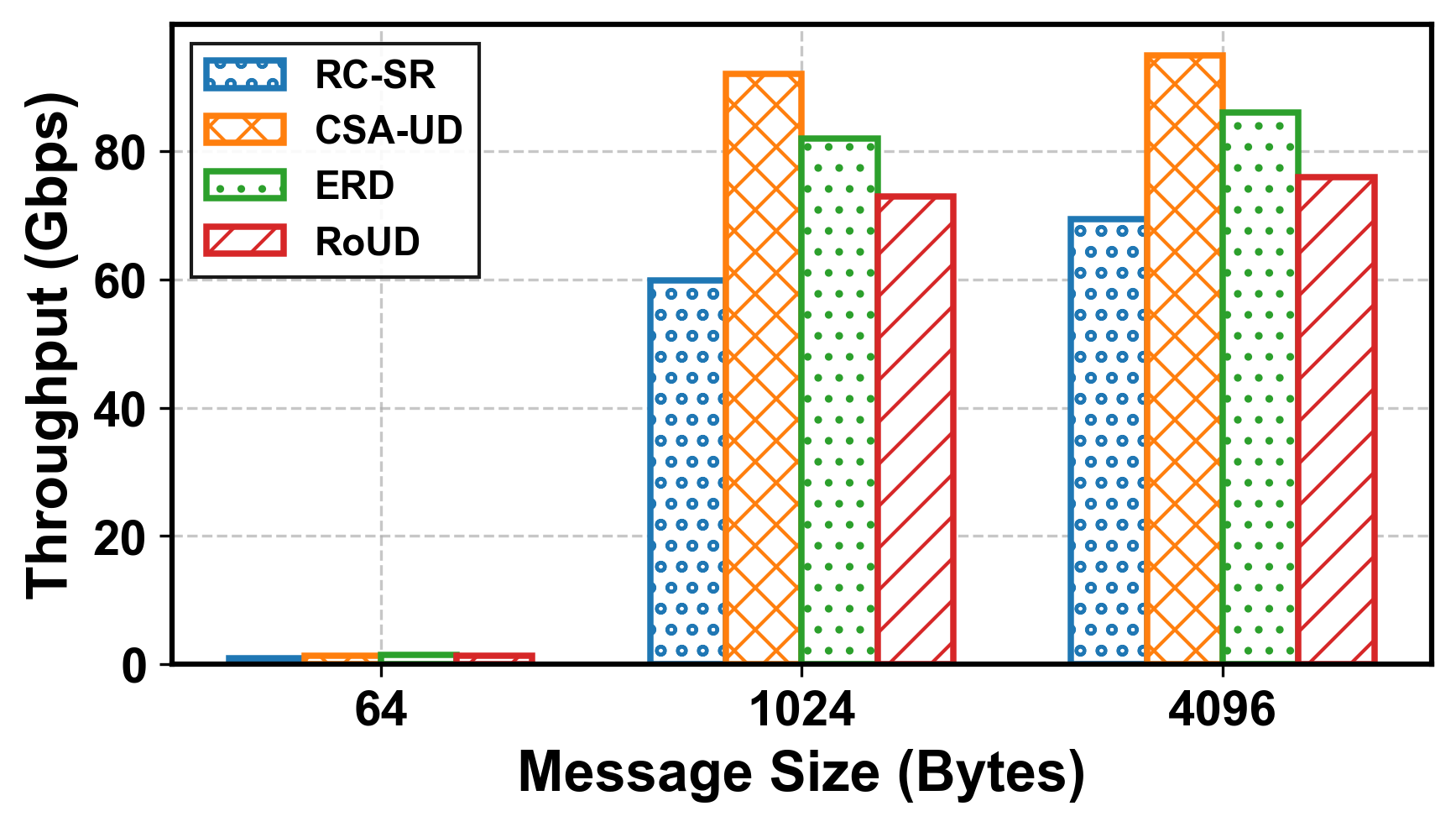}%
    \label{fig:schemes}
  }
  \caption{RNIC under different QP numbers and message sizes.}
  \label{fig:rnic-goodput}
\end{figure*}

\label{sec:packet-format}
\textbf{Experiment Results.}~To assess CSA-UD's performance under varying message sizes, we conduct experiments using QP numbers of 512, 1024, 2048, and 4096, while varying the message size among 64\,B, 1\,KB, and 4\,KB. The reported results represent the average over multiple runs to reduce measurement variance. as shown in Fig.~\ref{fig:rnic-goodput}. All schemes achieve very low throughput at 64\,B due to the high per-message overhead relative to the payload size. However, at 1\,KB and 4\,KB, CSA-UD consistently achieves the highest throughput across all QP settings, improving throughput by 12\%--54\% over RC-SR, 9\%--26\% over RoUD, and 2\%--12\% over ERD. The advantage grows with more QPs since RC-SR incurs heavy connection state and SACK-recovery overhead, RoUD retains inefficient SR-style recovery on UD, and ERD's fixed bitmap scanning ignores synchronization semantics, causing mistimed recovery and ineffective retransmissions. In contrast, CSA-UD uses communication-semantic-aware dynamic bitmap scanning to balance recovery efficiency and timeliness, and bitmap-guided lightweight reordering to sustain near-peak throughput across diverse configurations.

\textbf{Comparison algorithms.}~We compare CSA-UD with existing RDMA reliability  mechanisms, including ERD and RoUD (both UD-based), and RC with Selective Repeat (SR) for loss recovery. The evaluation focuses on the average and 99th percentile FCT under network loads ranging from 10\% to 100\%. RoUD uses a SACK-based Selective Repeat mechanism that maintains an expected PSN and infers losses based on packet arrival order. In contrast, ERD and CSA-UD adopt timer-based periodic loss detection and operate in a connectionless manner. To ensure fairness, we configure three ECMP paths per flow for both ERD and CSA-UD, consistent with the best-performing setup in ERD's original design. 
\subsection{Simulation Setup}

\textbf{Simulation Topology.}~We adopt a leaf-spine topology with 16 spine switches, 16 leaf switches, and 256 servers, where each spine switch connects to all leaf switches, forming a full bipartite graph. All links operate at 100 Gbps with per-hop latency in the microsecond range.

\textbf{Workload.}~We evaluate CSA-UD under an all-to-all collective  communication traffic load  to reflect typical patterns in distributed AI training. Each server sends data to all other nodes in every iteration, simulating full gradient exchanges. Flow generation adopts an on-off pattern, representing bursty and synchronized transmissions during training steps. The MTU size for each packet is fixed at 4\,KB.

\subsection{Simulation Results}
\begin{figure}[t]
  \centering
  \subfloat[Low Load]{%
    \includegraphics[width=0.48\linewidth]{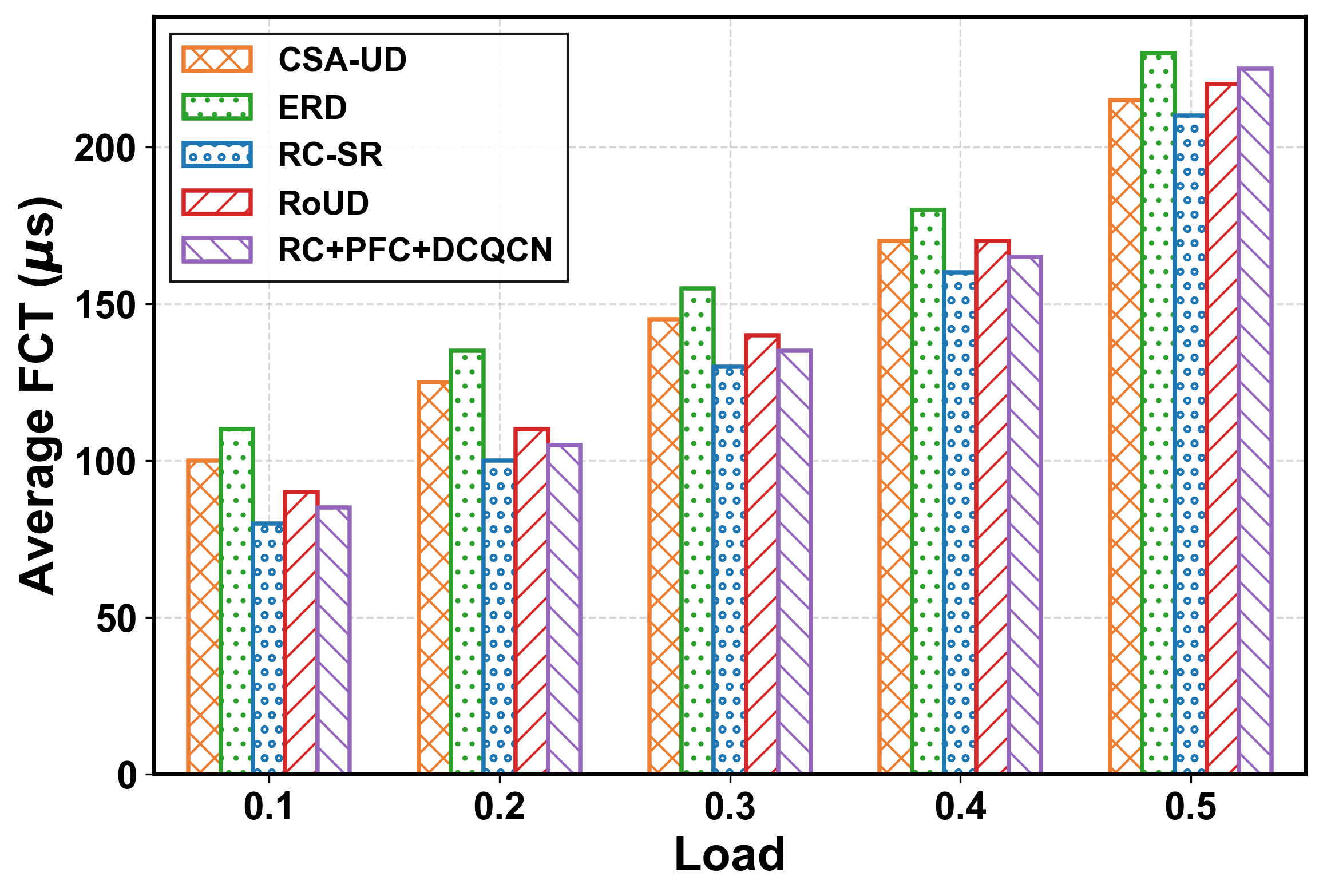}%
    \label{fig:avgfct-lowload}%
  }
  \hfill
  \subfloat[High Load]{%
    \includegraphics[width=0.48\linewidth]{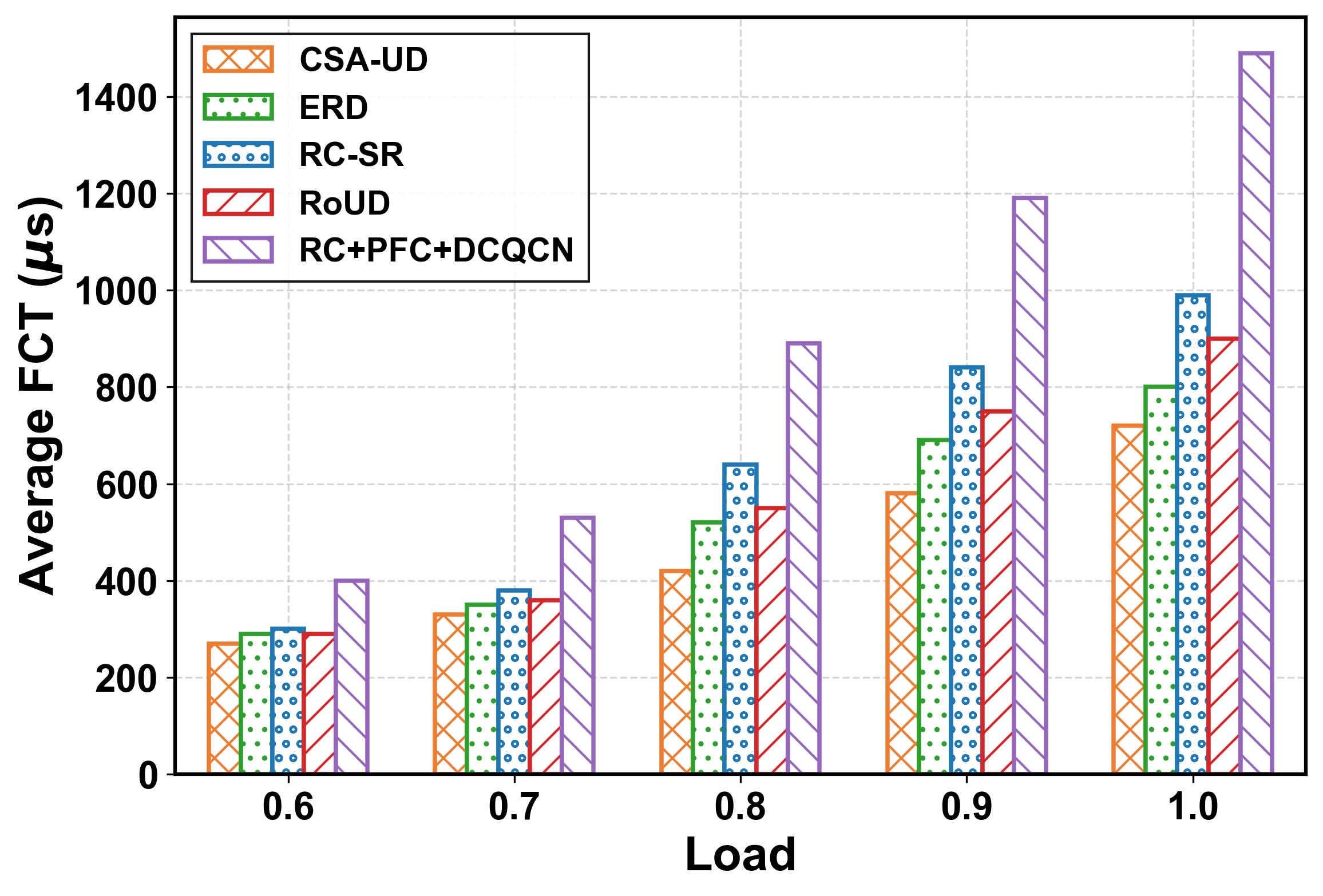}%
    \label{fig:avgfct-highload}%
  }
  \caption{Average FCT  of five RDMA transport mechanisms under (a)~low and (b)~high load conditions.}
  \label{fig:avgfct-comparison}
\end{figure}

\begin{figure}[t]
  \centering
  \subfloat[Low Load]{%
    \includegraphics[width=0.48\linewidth]{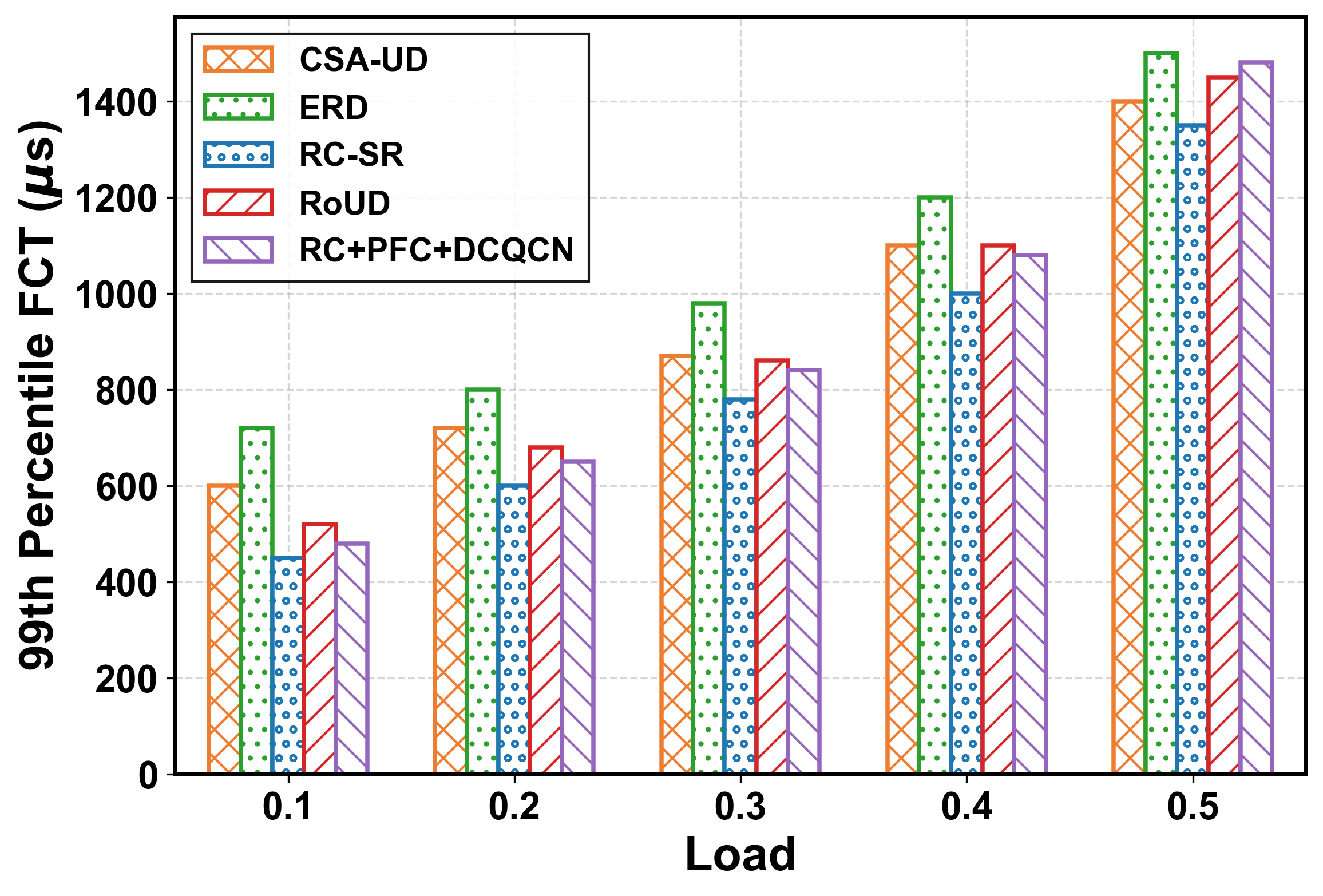}%
    \label{fig:fct-lowload}%
  }
  \hfill
  \subfloat[High Load]{%
    \includegraphics[width=0.48\linewidth]{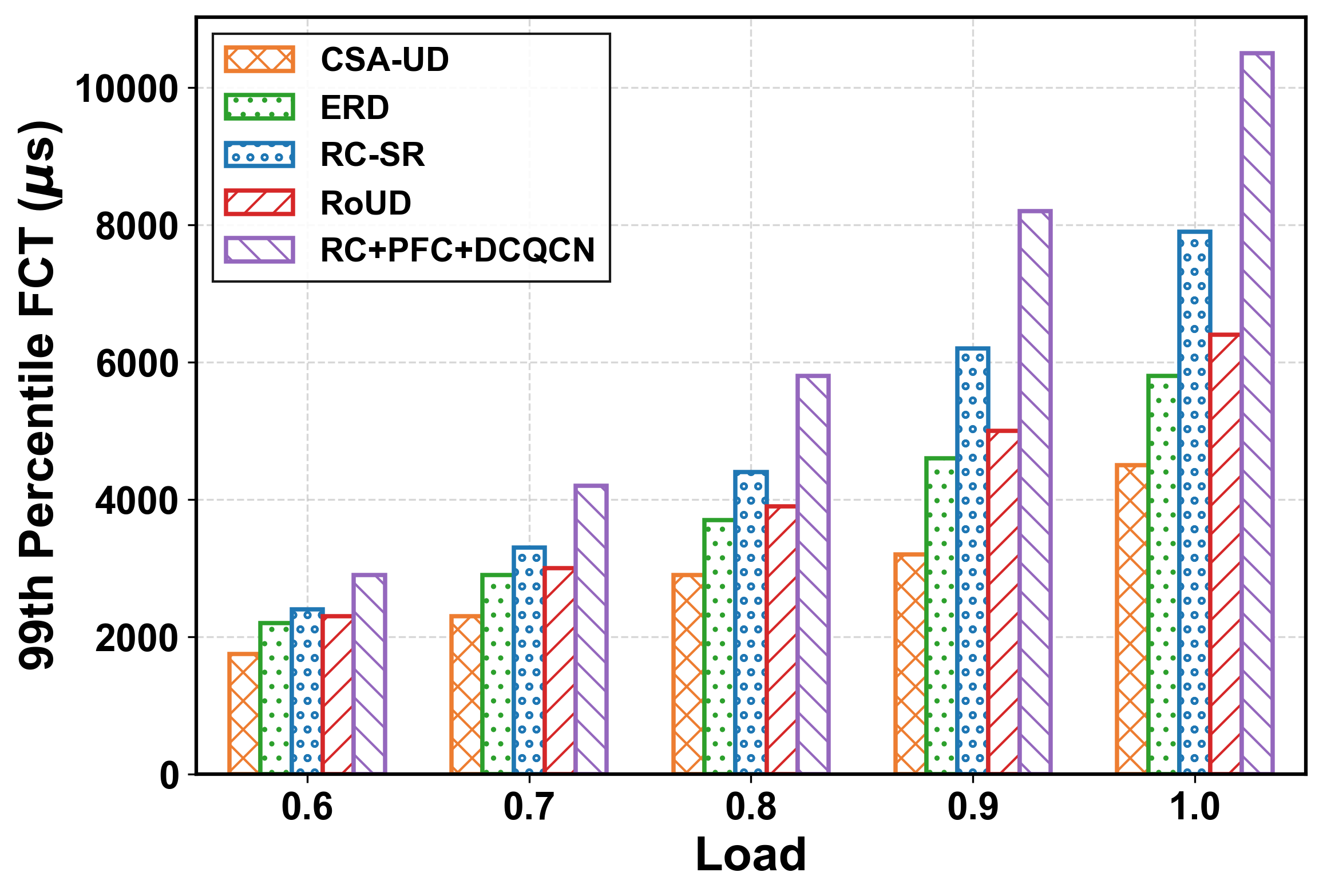}%
    \label{fig:fct-highload}%
  }
  \caption{99th percentile FCT of five evaluated mechanisms under (a)~low and (b)~high load conditions.}
  \label{fig:fct-99th}
\end{figure}
\begin{figure}[t]
  \centering
  \includegraphics[width=0.75\linewidth]{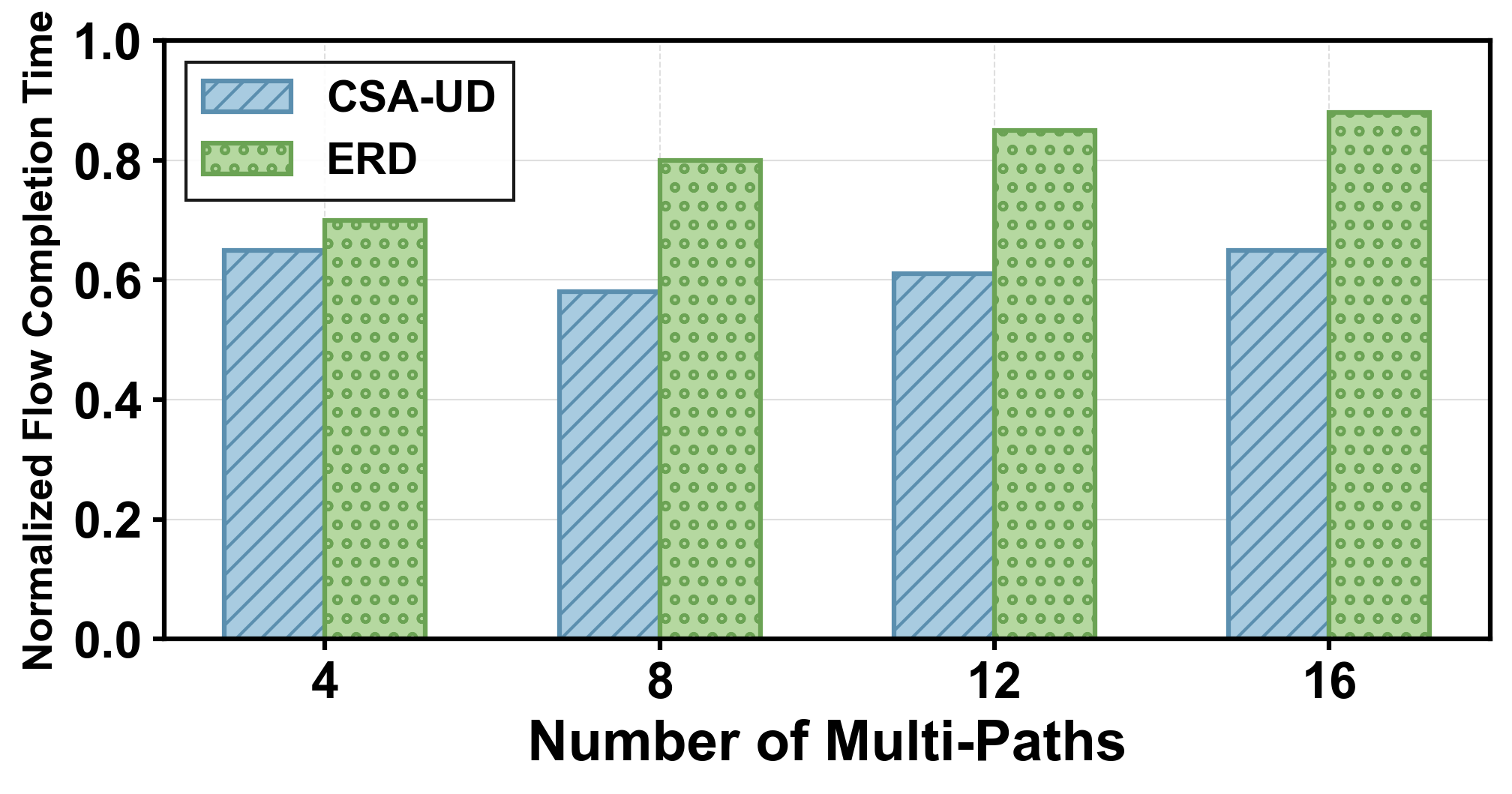}
  \caption{
    Impact of out-of-order packets  under multipath scenarios. 
  }
  \label{fig:reordering-nfct}
\end{figure}

\textbf{Flow Completion Time.}
\begin{figure}[t]
  \centering
  \subfloat[99th-Percentile FCT under Varying Load]{%
    \includegraphics[width=0.48\linewidth]{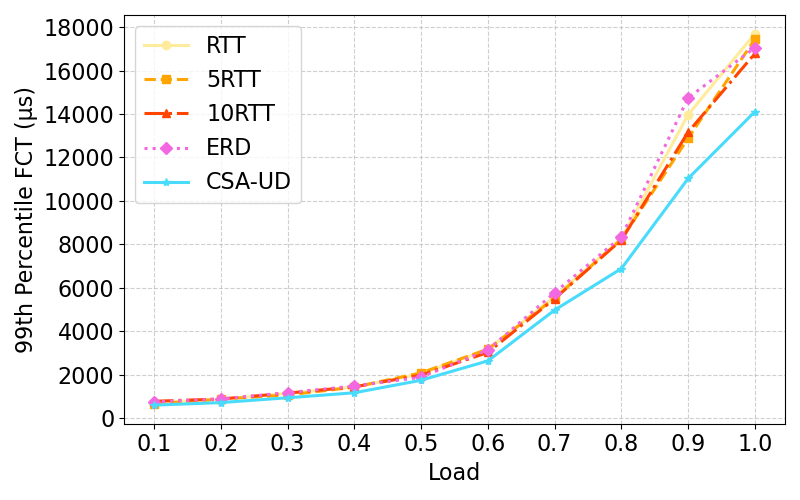}%
    \label{fig:micro-sim-a}
  }
  \hfill
  \subfloat[Retransmission Ratio under Varying Load]{%
    \includegraphics[width=0.48\linewidth]{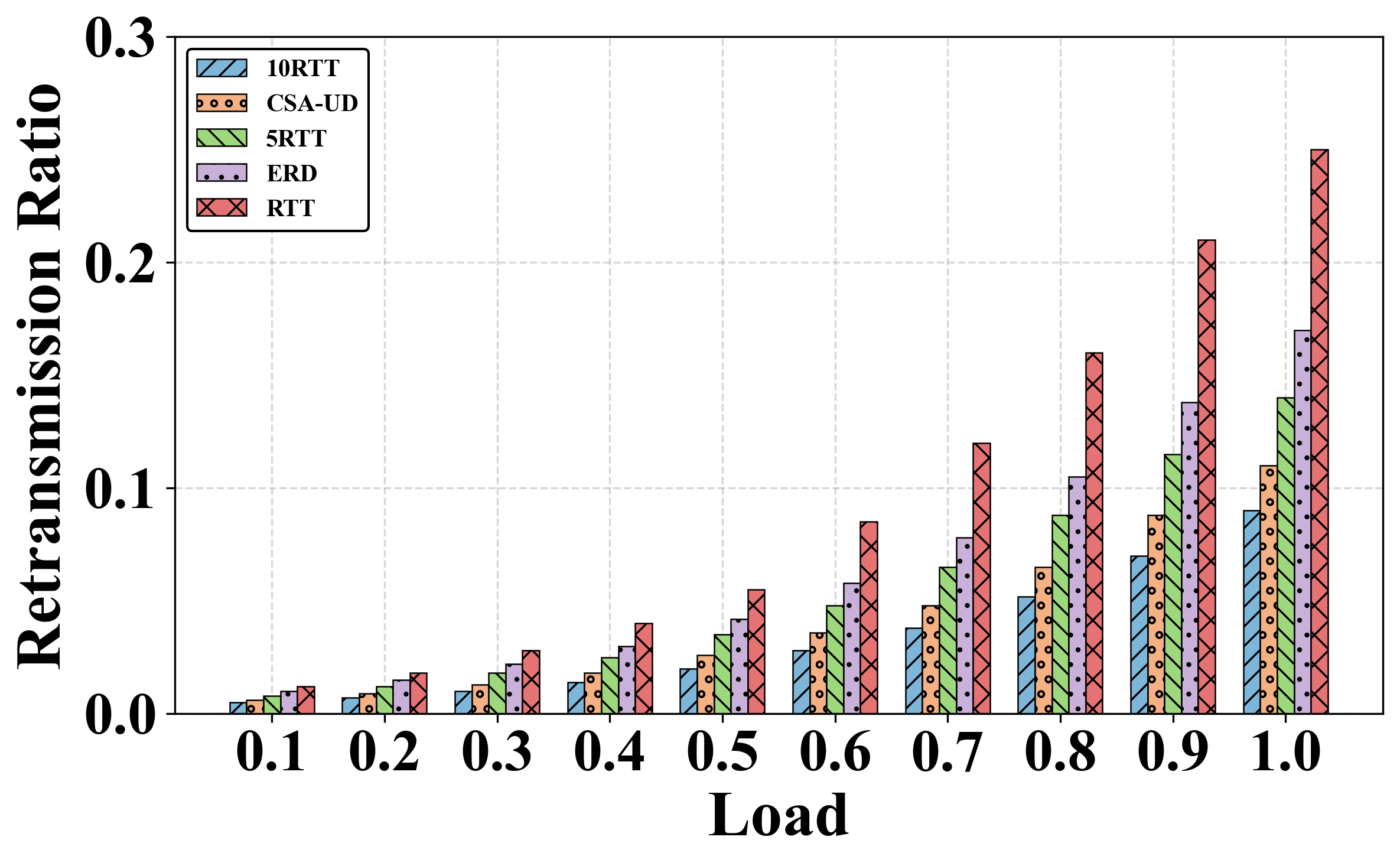}%
    \label{fig:micro-sim-b}
  }
  \caption{99th-percentile FCT and retransmission ratio under varying load.}
  \label{fig:micro-sim}
\end{figure}
Fig.~\ref{fig:avgfct-comparison} and Fig.~\ref{fig:fct-99th} present the average and 99th percentile FCT of five mechanisms under varying load levels.

Under low load (10\%--50\%), RC-SR achieves the lowest FCT among all schemes. Since congestion and packet losses are infrequent at low utilization, RC-SR's SACK-based recovery mechanism efficiently handles occasional losses without incurring reordering or userspace overhead. Nevertheless, this low-load advantage of RC-SR comes at the cost of poor scalability, as discussed below.  CSA-UD and other UD-based schemes introduce additional costs due to reassembly in userspace, leading to slightly higher FCT. Nevertheless, this low-load advantage of RC-SR comes at the cost of poor scalability, as discussed below. RC+PFC+DCQCN employs PFC-based lossless transport combined with DCQCN congestion control, representing the mainstream deployment in current RDMA data centers. While PFC eliminates packet loss by pausing upstream senders, it introduces severe HOL blocking that propagates congestion across the fabric. Even under low load, the overhead of ECN marking and rate adjustment already results in the highest FCT among all mechanisms.

Under high load (60\%--100\%), congestion and packet losses become increasingly prominent, and the performance landscape reverses. In RC+PFC+DCQCN, PFC pause frames cascade through the network, forming ``PFC storms'' that drastically degrade performance and trigger widespread HOL blocking. At 100\% workload, RC+PFC+DCQCN exhibits an average FCT of 1490~\textmu s---over 2$\times$ CSA-UD (720~\textmu s)---and a 99th percentile FCT of 10500~\textmu s, which is 2.3$\times$ CSA-UD (4500~\textmu s).

RC-SR relies on end-to-end SACK-driven recovery, which introduces additional control traffic and retransmissions under congestion and amplifies tail latency, especially with many concurrent flows. At 100\% load, RC-SR's average FCT reaches 990~\textmu s and its 99th percentile FCT reaches 7900~\textmu s, which is 1.76$\times$ that of CSA-UD. RoUD inherits the loss recovery logic of SR protocols but does not fully exploit the connectionless flexibility of UD transport. At 100\% load, RoUD exhibits a 99th percentile FCT of 6400~\textmu s---over 10\% higher than ERD (5800~\textmu s)---though it still outperforms both RC-SR and RC+PFC+DCQCN.

ERD adopts bitmap-driven loss detection, but its scanning interval $T$ remains fixed and does not adapt to congestion level, resulting in excessive NACKs and redundant retransmissions that further congest the network. ERD also performs multi-stage classification and sorting at the receiver, making it highly sensitive to packet reordering and increasing reassembly overhead, which limits ERD to a maximum of three paths in practice.
CSA-UD leverages a communication-semantic-aware design, combining adaptive NACK with efficient bitmap-guided packet reordering. Under high load (60\%--100\%), CSA-UD consistently achieves the lowest FCT among all five schemes. Compared with ERD, CSA-UD reduces the 99th percentile FCT by over 20\% across the entire high-load range, with the largest improvement of up to 30\% observed at 90\% load. Compared with the widely deployed RC+PFC+DCQCN, CSA-UD reduces the average FCT by up to 52\% and  99th percentile FCT by up to 57\% at full load, effectively eliminating the HOL blocking and PFC storm penalties inherent in lossless designs. Even against RC-SR, which excels under low load, CSA-UD achieves 27\% lower average FCT and 43\% lower 99th percentile FCT at full load. These results demonstrate CSA-UD's ability to suppress tail latency, reduce retransmission overhead, and maintain robust scalability in large-scale AI training workloads.

\textbf{Multipath Reordering Evaluation.}~As shown in Fig.~\ref{fig:reordering-nfct}, we evaluate the impact of multipath count on end-to-end FCT under a fixed network load of 70\%. We configure 4, 8, 12, and 16 equal-cost paths, and set the message size to 1024\,KB based on the goodput results to ensure stable measurements.

We compare CSA-UD and ERD using normalized FCT. CSA-UD consistently achieves lower FCT across all multipath settings. As the number of paths increases, both mechanisms benefit from load balancing and path diversity. However, ERD degrades due to intensified packet reordering: its receiver relies on group-based reassembly and in-order delivery, which become inefficient under severe out-of-order conditions, increasing reordering latency.

In contrast, CSA-UD scales better in multipath-dense environments. Its loss detection and reordering leverage dynamic bitmap scanning, remaining robust under complex path dynamics and maintaining stable FCT with minimal degradation. These results highlight the need for robust reordering support in UD-based RDMA protocols at high multipath counts.

\textbf{Impact of Adaptive Scan Interval.}~To evaluate adaptive scan scheduling, we compare CSA-UD with fixed-interval baselines (1\,RTT, 5\,RTT, 10\,RTT) and ERD under increasing network load.

Fig.~\ref{fig:micro-sim}(a) shows that CSA-UD achieves the lowest 99th-percentile FCT across the entire load range, reducing tail latency by up to 25.8\% compared to ERD at high load.

Fig.~\ref{fig:micro-sim}(b) reports the retransmission ratio. CSA-UD does not always minimize retransmissions, but avoids overly aggressive scanning (1\,RTT) that inflates redundant retransmissions and overly conservative scanning (10\,RTT) that delays critical loss recovery. By dynamically balancing timely loss detection and redundant retransmission avoidance, CSA-UD achieves a favorable trade-off. This is important for distributed large-model training: increasing scan frequency near completion accelerates recovery on the critical path, while conservative scanning early in the transfer suppresses repeated retransmissions and network overhead.

Notably, these gains are achieved solely by adapting the scan interval, without modifying retransmission or congestion-control logic.

\section{Related Work}

Recent efforts in both academia and industry have aimed to improve QP scalability and congestion handling in RDMA-based AI systems.

Solutions like StaR~\cite{star-icnp21} and SRNIC reduce RNIC state through state separation and cacheless designs, while SRD~\cite{aws-srd-blog} introduces reliable datagram transport via Nitro DPUs. Though scalable, these approaches require hardware changes or DPU deployment, increasing cost and complexity. In contrast, ERD achieves  deployability without RNIC modifications.

Software approaches such as RoUD and XRC reduce QP counts through UD transport and shared contexts, but struggle under large-scale topologies. ERD improves on this by enabling multipath delivery over few QPs to reduce hotspots.

Multipath RDMA techniques like MP-RDMA~\cite{mprdma-nsdi18}, PLB~\cite{plb-apnet22}  and MPTD~\cite{mptd-cluster24} offer RTT-aware load balancing, but often worsen QP scalability due to virtual QP overhead. MPRUD~\cite{udmp-ictc22} uses UD QPs for multipath but suffers from interference and head-of-line blocking at scale.

\section{CONCLUSION}
This paper presents CSA-UD, a scalable and highly efficient communication-semantic-aware RDMA loss recovery mechanism for large-scale AI training. 
By replacing the traditional RC mode with the UD mode, CSA-UD overcomes QP scalability bottlenecks in large-scale training. 
It dynamically adjusts bitmap scanning intervals based on the on-off nature of training traffic and adopts a page-table-inspired buffer segmentation scheme to significantly reduce reordering overhead under multipath delivery. 
Extensive ns-3 simulations show that CSA-UD reduces the 99th percentile FCT by over 30\% under All-to-All workloads compared to the latest UD-based design, ERD. 
In goodput tests at 70\% load and 1024\,KB message size, CSA-UD achieves 90\,Gbps goodput, surpassing ERD’s 78\,Gbps. 
Moreover, CSA-UD consistently maintains lower normalized FCT than ERD under all multipath conditions, while ERD’s NFCT degrades from just 4 paths, demonstrating CSA-UD’s superior robustness and reordering efficiency.
\section{Acknowledgment}
This work was supported by the National Natural Science Foundation of China under Grant 62472219 and Grant 62132007, the Natural Science Foundation of Jiangsu Province under Grant BK20242038, and the China Environment for Network Innovations (CENI).

\bibliographystyle{IEEEtran}
\bibliography{references}
\end{document}